\documentclass[useAMS,usenatbib,usegraphicx,epsfig]{aa}
\usepackage{txfonts,graphicx,epsfig}

\def\eps@scaling{.95}
\def\epsscale#1{\gdef\eps@scaling{#1}}
\def\plotone#1{\centering \leavevmode
\epsfxsize=\eps@scaling\columnwidth \epsfbox{#1}}

\def\micron{\mu {\rm m}}

\begin{document}

\title{Radiative transfer models of non-spherical prestellar cores}
\titlerunning{Radiative transfer models of non-spherical prestellar cores}

\author{D.~Stamatellos \inst{1}, A.P.~Whitworth \inst{1}, 
P.~Andr\'e \inst{2},\and D.~Ward-Thompson \inst{1}}

\authorrunning{D.~Stamatellos et al.}

\offprints{D.~Stamatellos\\ \email{D.Stamatellos@astro.cf.ac.uk}}

\def\lpacket{$L$-packet }
\def\lpackets{$L$-packets }

\institute{Department of Physics \& Astronomy, Cardiff University, 
        PO Box 913, 5 The Parade, Cardiff CF24 3YB, Wales, UK
\and
CEA, DSM, DAPNIA, Service d' Astrophysique, C.E.  Saclay, 
F-91191 Gif-sur-Yvette Cedex, France}
       
\date{Received ..., 2003; accepted ... , 2003}

\abstract{

We present 2D Monte Carlo radiative transfer simulations of 
prestellar cores. We consider two types of asymmetry: {\it 
disk-like} asymmetry, in which the core is denser towards the 
equatorial plane than towards the poles; and {\it axial} asymmetry, 
in which the core is denser towards the south pole than the north 
pole. In both cases the degree of asymmetry is characterized by 
the ratio $e$ between the maximum optical depth from the centre 
of the core to its surface and the minimum optical depth from the 
centre of the core to its surface. We limit our treatment here 
to mild asymmetries with $e = 1.5\,$ and $2.5\,$. We consider both 
cores which are exposed directly to the interstellar radiation 
field and cores which are embedded inside molecular clouds.

\hspace{0.3cm} The SED of a core is essentially independent 
of the viewing angle, as long as the core is optically thin. However, 
the isophotal maps depend strongly on the viewing angle. Maps at 
wavelengths longer than the peak of the SED (e.g. 850 $\mu$m) 
essentially trace the column-density. This is because at long 
wavelengths the emissivity is only weakly dependent on temperature, 
and the range of temperature in a core is small (typically 
$T_{\rm max}/T_{\rm min} \la 2$). Thus, for instance, cores with 
disk-like asymmetry appear elongated when mapped at 850 
$\mu$m from close to the equatorial plane. However, at 
wavelengths near the peak of the SED (e.g. 200 $\mu$m), the 
emissivity is more strongly dependent on the temperature, and 
therefore, at particular viewing angles, there are 
characteristic features which reflect a more complicated 
convolution of the density and temperature fields within 
the core.

\hspace{0.3cm} These characteristic features are on scales $1/5$ to 
$1/3$ of the overall core size, and so high resolution observations are 
needed to observe them. They are also weaker if the core is embedded in a 
molecular cloud (because the range of temperature within the core 
is then smaller), and so high sensitivity is needed to detect them. 
{\it Herschel}, to be launched in 2007, will in principle provide 
the necessary resolution and sensitivity at 170 to 250 $\mu$m. 


}
 
\maketitle

\section{Introduction}

\hspace{1.2em}
Prestellar cores are condensations in molecular clouds that are 
either on the verge of collapse or already collapsing (e.g. Myers 
\& Benson 1983; Ward-Thompson, Andr\'e \& Kirk 2002). They represent 
the initial stage of star formation and their study is important 
because theoretical models of protostellar collapse suggest that the 
outcome is very sensitive to 
the initial conditions. Prestellar cores have been observed both in 
isolation and  in protoclusters. Isolated prestellar cores (e.g. 
L1544, L43 and L63; Ward-Thompson, Motte \& Andr\'e 1999) have extents 
$\ga 1.5\times 10^4$ AU and  masses 
$0.5-35~{\rm M}_{\sun}$ (see also Andr\'{e}, Ward-Thompson, \& 
Barsony 2000). On the other hand, prestellar cores in protoclusters 
(e.g. in $\rho$ Oph and NGC2068/2071) are generally smaller, with 
extents $\sim 2-4 \times 10^3$ AU and masses $\sim 0.05-3~{\rm M}_{\sun}$ 
(Motte, Andr\'e  \& Neri 1998, Motte et al. 2001).

Many authors have modelled prestellar cores with Bonnor-Ebert 
(BE) spheres, i.e. equilibrium isothermal spheres in which 
self-gravity is balanced by gas pressure (Ebert 1955, Bonnor 
1956). For example, Barnard 68 has been modelled in this way by 
Alves, Lada \& Lada (2001). 

However, it is evident from 850 $\mu$m continuum maps of 
prestellar cores, which essentially trace the column-density 
through a core, that prestellar cores are not usually 
spherically symmetric (e.g. Motte et al., 1998; Ward-Thompson 
et al., 1999; Kirk et al., 2004). Indeed, statistical analyses 
of the projected shapes of a large sample of cores (Jijina, 
Myers \& Adams, 1999) suggest that prestellar cores do not 
even have spheroidal symmetry and are better represented by 
triaxial ellipsoids (Jones, Basu \& Dubinski, 2001; Goodwin, 
Ward-Thompson \& Whitworth, 2002).

This is not surprising, given the highly turbulent nature of 
star-forming molecular clouds, and the short time-scale on which 
star formation occurs (e.g. Elmegreen, 2000). Normally, prestellar 
cores are formed -- and then either collapse or disperse -- so 
rapidly that they do not have time to relax towards equilibrium 
structures. Even in more quiescent environments where cores can 
evolve quasi-statically, the combination of magnetic and rotational 
stresses is likely to produce significant departures from spherical 
symmetry.

For instance, SPH simulations of isothermal, turbulent, molecular 
clouds by Ballesteros-Paredes, Klessen \& Vazquez-Semadeni (2003) 
show (a) that most of the cores that form are transient and 
non-spherical; and (b) that, despite this fact, the column density 
can, more often than not, be adequately fitted with the Bonnor-Ebert 
profile, although the parameters of the fit depend on the observer's 
viewing angle. Likewise, the FD simulations of magnetic, 
isothermal, turbulent, molecular clouds reported by Gammie et al. 
(2003) produce transient, triaxial cores.

Similarly, evolutionary models of individual prestellar cores predict the 
formation of flattened (oblate spheroidal) structures, either 
due to rotation (e.g. Matsumoto, Hanawa \& Nakamura 1997), or 
due to flattening along the lines of a bipolar magnetic 
field (e.g. Ciolek \& Mouschovias, 1994). Other models invoke 
a toroidal magnetic field to create prolate equilibrium cores 
(e.g. Fiege \& Pudritz, 2000). Triaxial structures can be 
generated with a suitable combination of rotation and magnetic 
field.

It is therefore important to investigate, by means of radiative 
transfer modelling, how the intrinsic asymmetries inherent in 
the formation and evolution of a core might translate into 
observable asymmetries on continuum maps of cores. Previous 
continuum radiative transfer modelling of prestellar cores 
has examined non-embedded BE spheres (Evans et al., 2001, 
Young et al., 2003) and embedded BE spheres (Stamatellos \& 
Whitworth, 2003a), using 1D (spherically-symmetric) codes. 
Zucconi, Walmsley \& Galli (2001) have used an approximate, 
semi-analytic method to model non-embedded, magnetically flattened 
prestellar cores, in 2D. We have recently developed a Monte 
Carlo code for modelling continuum radiative transfer in 
arbitrary geometry, and with arbitrary accuracy. Preliminary 
results have been presented in Stamatellos \& Whitworth 
(2003b,c).

In this paper we develop continuum radiative transfer 
models of non-spherical cores. Since these models are 
intended to be exploratory, rather than definitive, we consider 
here only 2D models (i.e. we impose azimuthal symmetry so that 
in spherical polar co-ordinates $(r,\theta,\phi)$ there is no 
dependence on $\phi$).

Since star formation is a chaotic process, it is not 
sensible to appeal to numerical simulations for the detailed 
density field in a prestellar core. From both observations 
(e.g. Kirk et al., 2004) and simulations (e.g. Goodwin et al., 
2004), it is clear that each core has a unique distribution 
of gas in its outer envelope, and a unique radiation field 
incident on its boundary. Even if existing simulations are 
a good representation of the real dynamics of star formation, 
they are not presently able to reproduce particular sources, 
and therefore they can only be compared realistically with 
observations in a statistical sense.

We will be presenting SEDs and isophotal maps for prestellar 
cores formed in SPH simulations of star formation in turbulent 
molecular clouds in a subsequent paper (Stamatellos, Goodwin 
\& Whitworth, in preparation). However, for interpreting 
observations of individual cores, it is more appropriate to 
generate SEDs and isophotal maps using simply parametrized 
models which capture generically the different features we 
might hope to detect.

In this regard, we have been guided by the 
observations, which indicate that prestellar cores have 
approximately uniform density in their central regions, and 
the density then falls off in the envelope. If the density 
in the envelope is fitted with a power law, $n(r) \propto 
r^{-\eta}\,$, then $\eta \sim 2\,-\,4 \,$. Here 
$\eta \sim 2$ is characteristic of more extended prestellar 
cores in dispersed star formation regions (e.g. L1544, 
L63 and L43), whereas $\eta \sim 4$ is characteristic of more 
compact cores in protoclusters (e.g. $\rho$ Oph and NGC2068/2071). 

These features are conveniently represented by a Plummer-like 
density profile (Plummer, 1915),
\begin{equation}
n(r) = n_0\,\frac{1}{ \left[ 1 + \left( \frac{r}{r_0} 
\right)^2 \right]^{\eta/2} } \,,
\end{equation}
where $n_0$ is the density at the centre of the core, and 
$r_0$ is the extent of the region in which the density is 
approximately uniform. 

The Plummer-like density profile is ad hoc, but given the 
transient, non-hydrostatic nature of prestellar cores, and 
the coarseness of the observational constraints, this is 
unavoidable. It has the advantage of being simple, with only 
three free parameters. Uniquely amongst analytic models, it 
predicts lifetimes, accretion rates, collapse velocity fields, 
SEDs and isophotal maps which agree well with observation 
(Whitworth \& Ward-Thompson, 2001; Young et al., 2003). 
It also reproduces approximately the BE density profile, and the 
density profiles predicted by the ambipolar diffusion models 
of Ciolek \& Mouschovias (1994) and Ciolek \& Basu (2000). 
Tafalla et al. (2004) use a similar density profile to model the starless 
cores  L1498 and L1517B in Taurus-Auriga.

Furthermore, the Plummer-like profile 
can be modified easily 	to include azimuthally 
symmetric departures from spherical symmetry. We treat two 
types of asymmetry. In the first type (disk-like asymmetry), 
we construct flattened cores, using density profiles of the form
\begin{equation}
n(r,\theta) = n_0\,\frac{1 + A \left( \frac{r}{r_0} \right)^2 
{\rm sin}^p(\theta) }{ \left[ 1 + \left( \frac{r}{r_0} \right)^2 
\right]^{(\eta+2)/2} } \,.
\end{equation}
The parameter $A$ determines the equatorial-to-polar optical 
depth ratio $e\,$, i.e. the maximum optical depth from the centre 
to the surface of the core (which occurs at $\theta = 
90\degr$), divided by the minimum optical depth from the centre 
to the surface of the core (which occurs at $\theta = 0\degr$ 
and $\theta = 180\degr$). The parameter $p$ determines how rapidly 
the optical depth from the centre to the surface rises with 
increasing $\theta$, i.e. going from the north pole at $\theta 
= 0\degr$ to the equator at $\theta = 90\degr$.

In the second type (axial asymmetry), we construct cores which 
are denser towards the south pole ($\theta = 180\degr$) 
than the north pole ($\theta = 0\degr$), using density 
profiles of the form
\begin{equation}
n(r,\theta) = n_0\,\frac{1 + A \left( \frac{r}{r_0} \right)^2 
{\rm sin}^p(\theta/2) }{ \left[ 1 + \left( 
\frac{r}{r_0} \right)^2 \right]^{(\eta+2)/2} } \,.
\end{equation}
Here the parameter $A$ determines the south-to-north pole optical 
depth ratio $e\,$, i.e. the maximum optical depth from the centre 
to the surface of the core (which now occurs at $\theta = 180\degr$), 
divided by the minimum optical depth from the centre to the 
surface of the core (which still occurs at $\theta = 0\degr$). 
The parameter $p$ again determines how rapidly the optical depth 
from the centre to the surface rises with increasing $\theta$, but 
now going from the north pole at $\theta = 0\degr$ all the way 
round to the south pole at $\theta = 180\degr$.

In all models the core has a spherical boundary at radius 
$R_{\rm core}$.

For the purpose of this paper, and in order 
to isolate a manageable parameter space, we fix $n_0 = 10^6\,
{\rm cm}^{-3}$, $r_0 = 2 \times 10^3\,{\rm AU}$, $\eta = 2$, 
and $R_{\rm core} = 2 \times 10^4\,{\rm AU}$. These are 
typical values for isolated cores. We can then explore the 
effect of varying $A$ and $p$, or equivalently $e$ and $p$.

In Section 2, we outline the basic principles underlying our 
Monte Carlo radiative transfer code. In Section 3, we present 
results obtained for cores having disk-like asymmetry; we treat 
both non-embedded cores and cores embedded in molecular clouds. 
In Section 4, we present results for cores having axial asymmetry. 
In Section 5, we summarize our results.

\section{Numerical method \& initial system setup}

\label{sec:phaethon.info}

Our method (Stamatellos \& Whitworth 2003a) is similar to that 
developed by  Wolf, Henning \& Stecklum (1999) and Bjorkman \& 
Wood (2001), and is based on the fundamental principle of Monte 
Carlo methods, according to which we can sample a physical quantity 
from a probability distribution using random numbers. We represent 
the radiation field of a source (star or background radiation) by 
a large number of monochromatic luminosity packets ($L$-packets). 
These $L$-packets are injected into the system and interact 
stochastically with it. If an $L$-packet is absorbed its energy is 
added to the local cell and raises the local temperature. To ensure 
radiative equilibrium  the $L$-packet is re-emitted immediately with 
a new frequency chosen from the difference between the local cell 
emissivity before and after the absorption of the packet (Bjorkman 
\& Wood 2001). This method conserves energy exactly, accounts for 
the diffuse radiation field, and its 3-dimensional nature makes it 
attractive for application to a variety of systems. The code ({\sc 
Phaethon}) has been thoroughly tested using the thermodynamic 
equilibrium test (Stamatellos  \& Whitworth 2003a), and also against 
the benchmark calculations defined by Ivezic et al. (1997) and by 
Bjorkman \& Wood (2001).

\begin{figure}
\centerline{
\includegraphics[width=7.3cm]{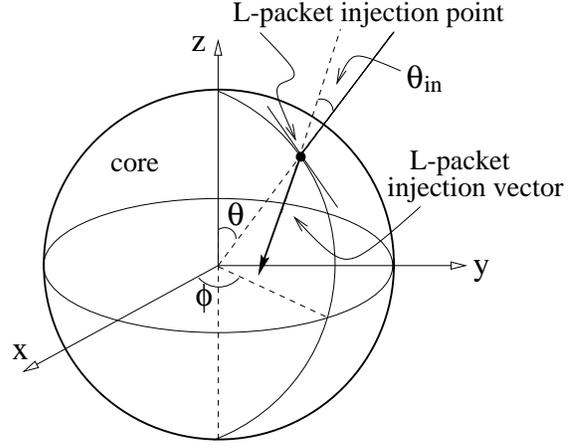}}
\caption{L-packet injection into the core: The packet is injected 
from a random point ($\theta$,$\phi$) on the surface of the sphere 
at such an angle ($\theta_{\rm in}$,$\phi_{\rm in}$) as to imitate 
an isotropic radiation field. The frequency of the injected packet 
is chosen from the Black (1994) radiation field.}
\label{fig_asym.injection}
\end{figure}

The code used here is adapted for the study of cores having azimuthal 
symmetry. The core itself is divided into a number of cells by spherical 
and conical surfaces. The spherical surfaces are evenly spaced in radius, 
and there are typically 50-100 of them. The conical surfaces are evenly 
spaced in polar angle, and there are typically 10-20 of them. Hence the 
core is divided into 500-2000 cells. The specific number of cells used 
is chosen so that the density and temperature differences between 
adjacent cells are small.

The \lpackets are injected from  the outside of the core with injection 
point and injection direction chosen to mimic an isotropic radiation 
field incident on the core (see Fig.~\ref{fig_asym.injection}). We first 
generate the \lpacket injection point on the surface of the core using 
random numbers $\mathcal{R}_1,\mathcal{R}_2~\in~[0,1]$, 
\begin{eqnarray}
r & = & R_{\rm core}\,, \\
\theta & = & \cos^{-1}(1-2\mathcal{R}_1)\;, \\
\phi & = & 2\pi\mathcal{R}_2\;;
\end {eqnarray} 
and then the injection direction (Fig.~\ref{fig_asym.injection}), using 
random numbers $\mathcal{R}_3,\mathcal{R}_4~\in~[0,1]$,
\begin {eqnarray}
\theta_{\rm in} & = & \cos^{-1}\left(\mathcal{R}_3^{1/2}\right)\;, \\
\phi_{\rm in} & = &  2\pi\mathcal{R}_4\;.
\end {eqnarray} 
Here $\theta_{\rm in}$ is the angle between the normal vector to the 
tangent plane at the point of entry and the packet injection vector, 
and $\phi_{\rm in}$ is the polar angle defined on the tangent plane.

We further assume that the radiation field incident on the core is 
the Black (1994) interstellar radiation field (hereafter BISRF),
which consists of radiation from giant stars and dwarfs, thermal 
emission from dust grains, mid-infrared emission from transiently 
heated small grains, and the cosmic background radiation. This is 
a good approximation to the radiation field in the solar neighbourhood 
but it is not always a good choice when studying prestellar cores, since 
in many cases the immediate core environment plays an important role in 
determining the radiation field incident on the core. In this work, we 
simulate the effect of the immediate core environment by embedding the 
core in a molecular cloud which modulates the radiation field incident 
on the core. Another option is to estimate the effective radiation field 
incident on an embedded core by observing it directly (Andr\'e et al. 2003).

The dust composition (and therefore the dust opacity) in prestellar cores 
is uncertain, but in such cold and dense conditions, dust particles are
expected to coagulate and accrete ice. As in our previous study of
prestellar cores (see Stamatellos \& Whitworth 2003a), we use the 
Ossenkopf \& Henning (1994) opacities for a standard MRN interstellar
grain mixture (53\% silicate and 47\% graphite) that has coagulated and 
accreted thin ice mantles over a period of $10^5$ yr at a density of 
$10^6\; {\rm cm}^{-3}$.

\section{Disk-like asymmetry}
\label{sec:flattened.cores}

\subsection{The model} 

The density profile in model cores with disk-like asymmetry is 
given by
\begin{equation}
n(r,\theta) = 10^6\,{\rm cm}^{-3}\;\,\frac{1 + A \left( \frac{r}
{2000\,{\rm AU}} \right)^2 {\rm sin}^p(\theta) }{ \left[ 1 + 
\left( \frac{r}{2000\,{\rm AU}} \right)^2 \right]^2 } \,.
\end{equation}
Thus the density is approximately uniform in the centre, and 
falls off as $r^{-2}$ in the outer envelope. The core has a 
spherical boundary at radius $R_{\rm core} = 2 \times 10^4\,
{\rm AU}$. The degree of asymmetry is determined by $A$ and $p$, 
and the values we have treated are given in Table 1, along with 
\begin{equation}
e = \frac{\tau_{\rm V}(\theta = 90\degr)}
{\tau_{\rm V}(\theta = 0\degr)} \,,
\end{equation}
$M_{\rm core}$ and $\tau_{\rm V}(\theta = 0^{\rm o})$. 
$e$ is the equatorial-to-polar optical depth ratio, i.e. the 
ratio of the maximum optical depth from the centre to 
the surface of the core (which occurs at $\theta = 90\degr$) 
to the minimum optical depth from the centre to the surface of 
the core (which occurs at $\theta = 0\degr$). Figure 2 displays 
isodensity contours on the $x=0$ plane for all the cases in Table 1.

\begin{table}
\begin{center}
\caption{Model parameters for cores with disk-like asymmetry}
\begin{tabular}{@{}cccccc} \hline
Model ID & $A$ & $p$ & $e$ & $M_{\rm core}/M_\odot$ & 
$\tau_{\rm V}(\theta = 0\degr)$ \\ \hline\hline
1.1 & 28 & 4 & 1.5 & 2.0 & 94 \\
1.2 & 81 & 4 & 2.5 & 5.1 & 94 \\
1.3 & 28 & 1 & 1.5 & 5.0 & 94 \\
1.4 & 81 & 1 & 2.5 & 7.3 & 94 \\ \hline\hline
\end{tabular}
\begin{list}{}{}
\item[$e\;\;$:] Equatorial-to-polar optical depth ratio
\item[$M_{\rm core}\;\;$:] Core mass
\item[$\tau_V(\theta = 0\degr)\;\;$:] Visual optical depth from the 
centre to the surface of the core along the pole ($\theta=0\degr$).
\end{list}
\label{tab:run_params2}
\end{center}
\end{table}

\subsection{Results: core temperatures, SEDs and images}

In non-embedded cores, the dust temperature drops from around 17~K 
at the edge of the core to 7~K at the centre of the core, as previous 
studies have already indicated (Zucconi et al. 2001; Evans et al. 2001; 
Stamatellos \& Whitworth 2003a). We also find that the dust temperature 
inside cores with disk-like asymmetry is $\theta$ dependent (see 
Figs.~\ref{fig_temp.asyma}), similar to the results of Zucconi et al. 
(2001). As expected, the equator of the core is colder than the poles. 
The difference in temperature between two points having the same distance 
$r$ from the centre of the core but with different polar angles $\theta$, 
is larger for the more asymmetric cores (i.e. those with larger $e$ 
and/or $p$; Figs.~\ref{fig_temp.asyma}). For example, at half 
the radius of the core ($10^4$~AU) the temperature difference 
between the  point at $\theta=0\degr$ (core pole) and the point at 
$\theta=90\degr$ (core equator), is $5-6$~K for the $p=4$ models 
(Figs.~\ref{fig_temp.asyma}a,~\ref{fig_temp.asyma}b) but only $\sim 2$~K 
for the $p=1$ models (Figs.~\ref{fig_temp.asyma}c,~\ref{fig_temp.asyma}d). 
This temperature difference will affect the appearance of the core at 
wavelengths shorter than or near the core peak emission, where the Planck 
function is strongly (exponentially) dependent on the temperature.

\begin{figure}[hp]

\centerline{
\includegraphics[width=5cm,angle=-90]{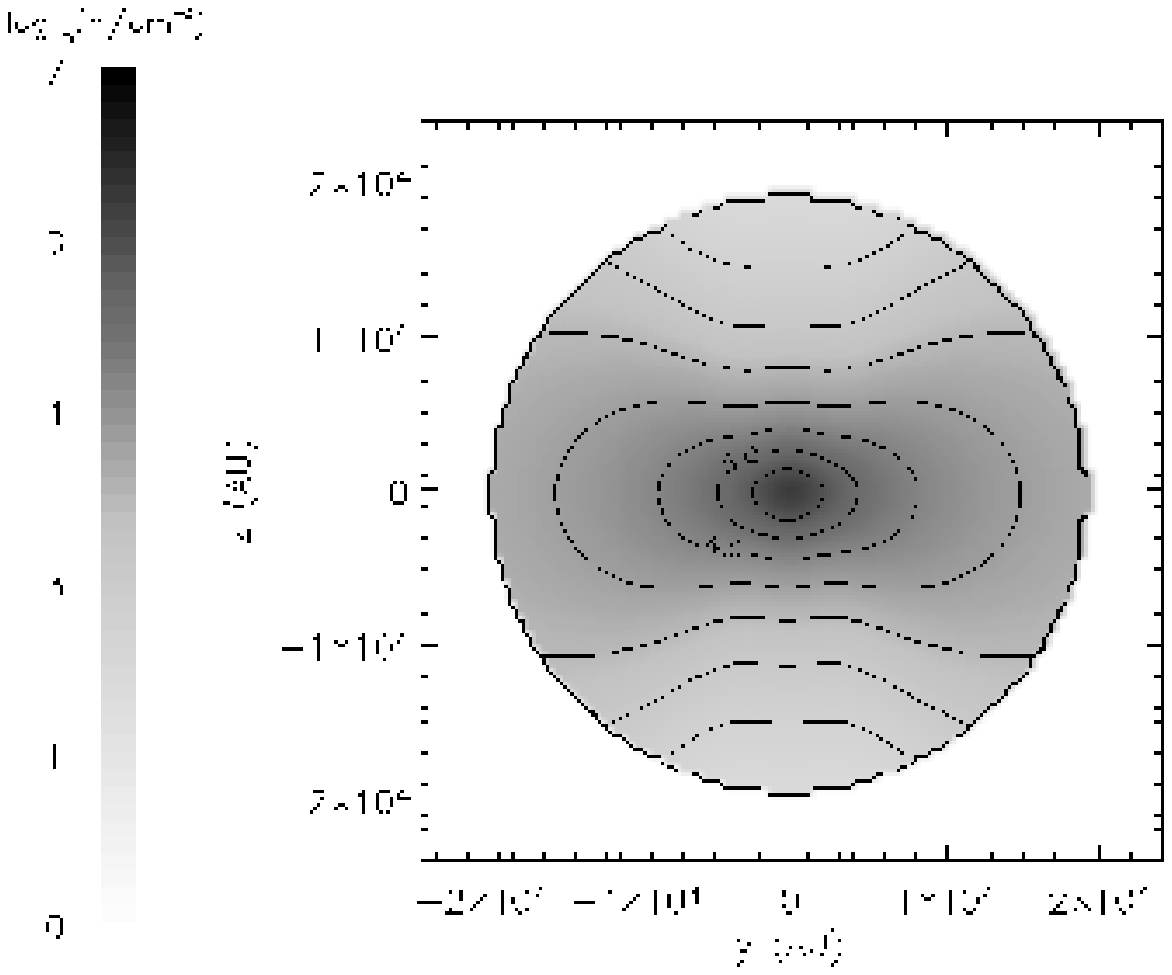}}
\vspace{-3cm}\hspace{1cm}{a}\vspace{+3cm}

\centerline{
\includegraphics[width=5cm,angle=-90]{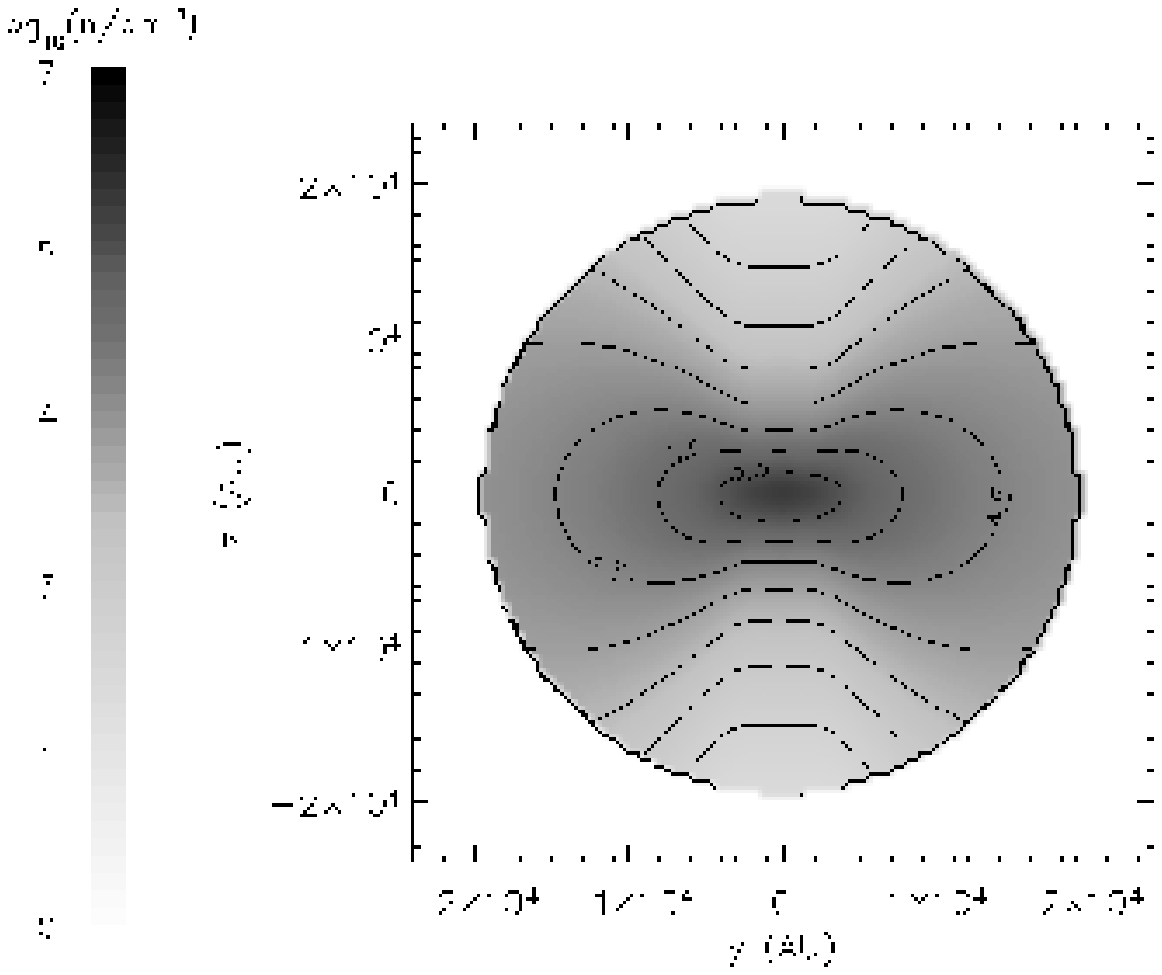}}
\vspace{-3cm}\hspace{1cm}{b}\vspace{+3cm}

\centerline{
\includegraphics[width=5cm,angle=-90]{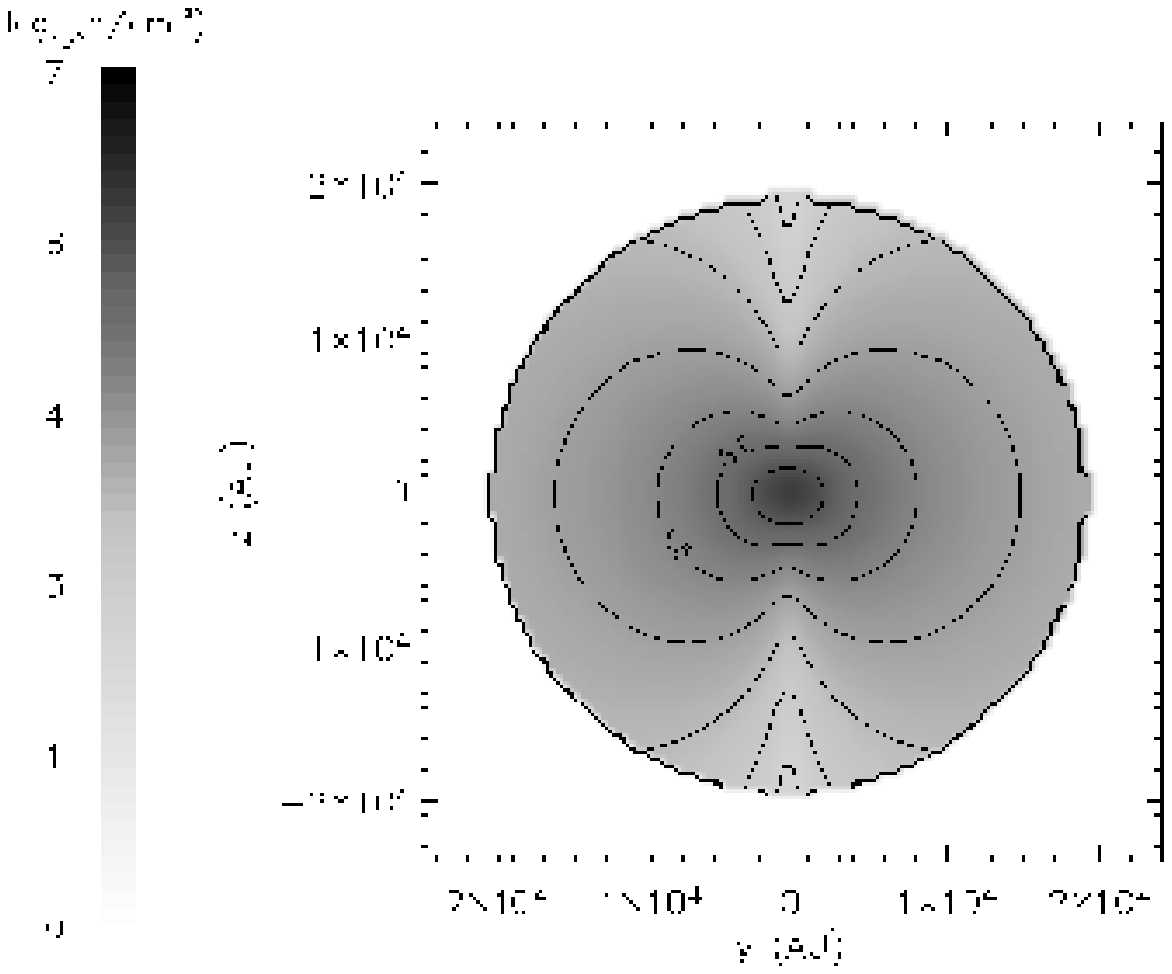}}
\vspace{-3cm}\hspace{1cm}{c}\vspace{+3cm}

\centerline{
\includegraphics[width=5cm,angle=-90]{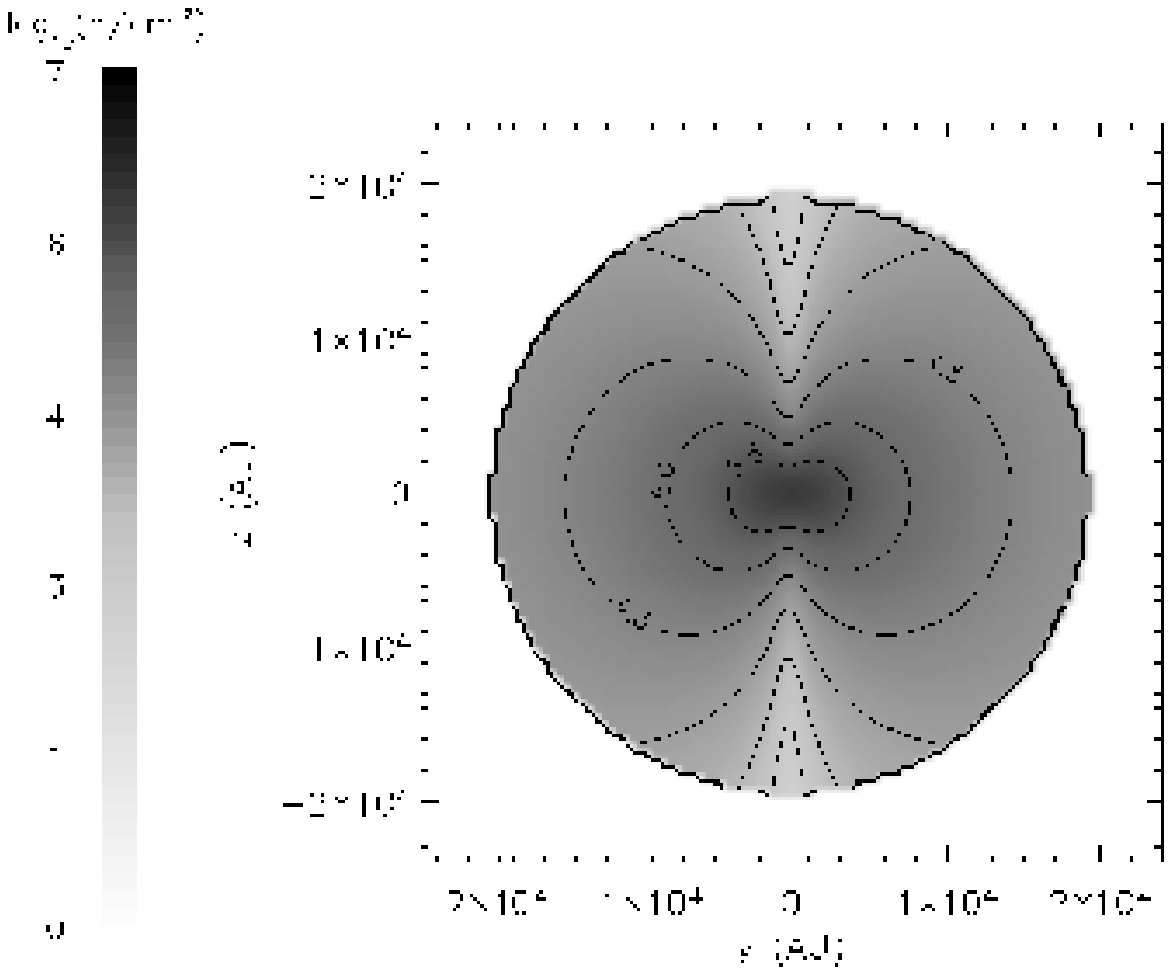}}
\vspace{-3cm}\hspace{1cm}{d}\vspace{+3cm}
\caption{Density distribution on the $x=0$ plane (a) for a flattened 
asymmetric core with equatorial-to-polar optical depth ratio $e=1.5$ 
and $p=4$ (model 1.1), (b) for a more flattened asymmetric core, with 
$e=2.5$ and  $p=4$ (model 1.2), (c) for a core with  $e=1.5$ and $p=1$ 
(model 1.3), and (d) for a more flattened core, with  $e=2.5$ and $p=1$ 
(model 1.4). We plot iso-density contours every $10^{0.5} {\rm cm^{-3}}$. 
The central contour corresponds to $n=10^{5.5} {\rm cm^{-3}}$.}
\label{fig_dens.asyma}
\label{fig_dens.asyma.1.5}
\label{fig_dens.asyma2.1.5}
\label{fig_dens.asyma.2.5}
\label{fig_dens.asyma2.2.5}
\end{figure}

\begin{figure}[!hp]
\centerline{
\includegraphics[width=5cm,angle=-90]{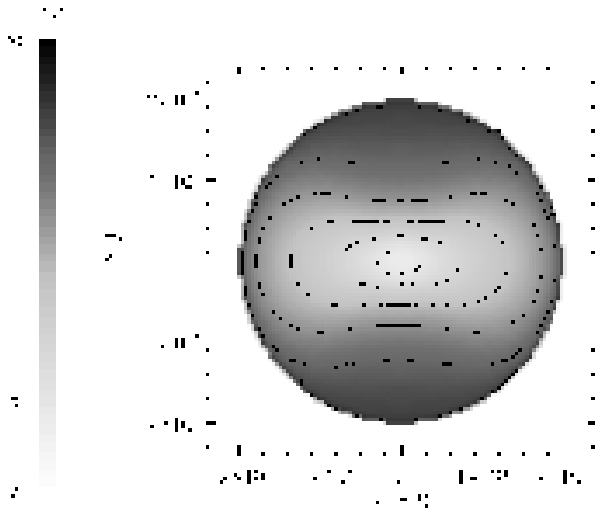}}
\vspace{-3cm}\hspace{1cm}{a}\vspace{+3cm}

\centerline{
\includegraphics[width=5cm,angle=-90]{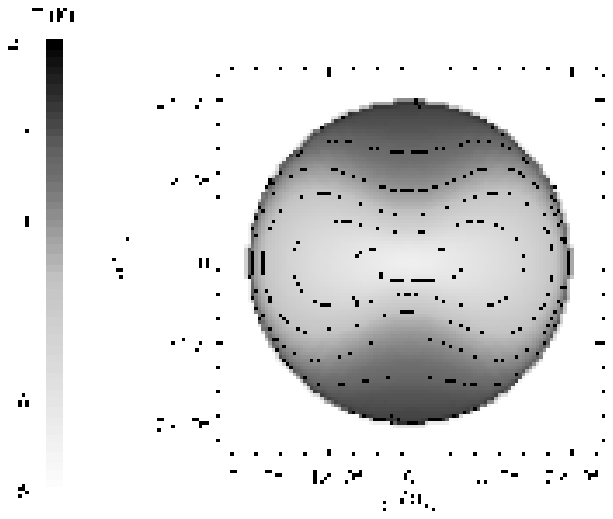}}
\vspace{-3cm}\hspace{1cm}{b}\vspace{+3cm}

\centerline{
\includegraphics[width=5cm,angle=-90]{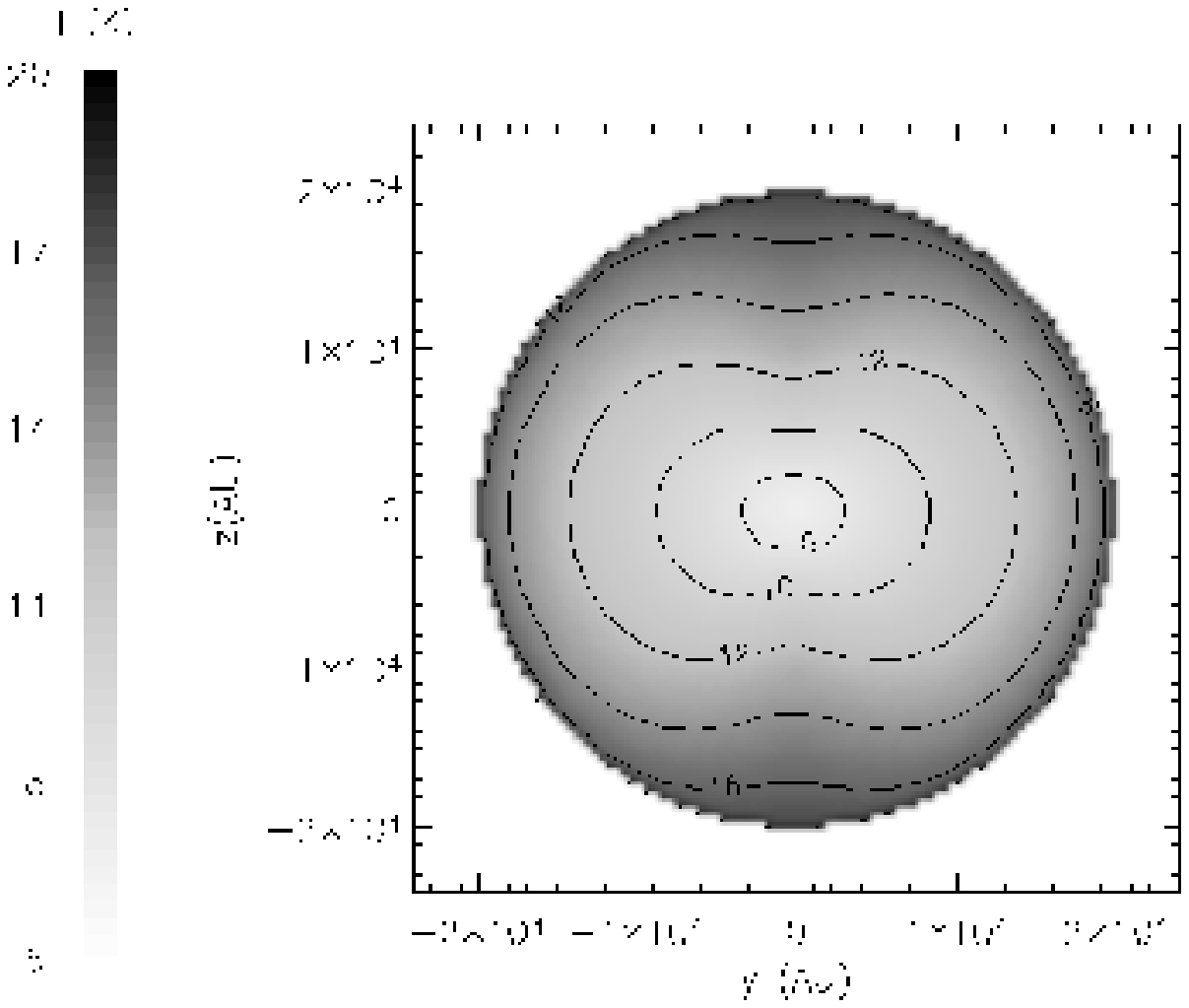}}
\vspace{-3cm}\hspace{1cm}{c}\vspace{+3cm}

\centerline{
\includegraphics[width=5cm,angle=-90]{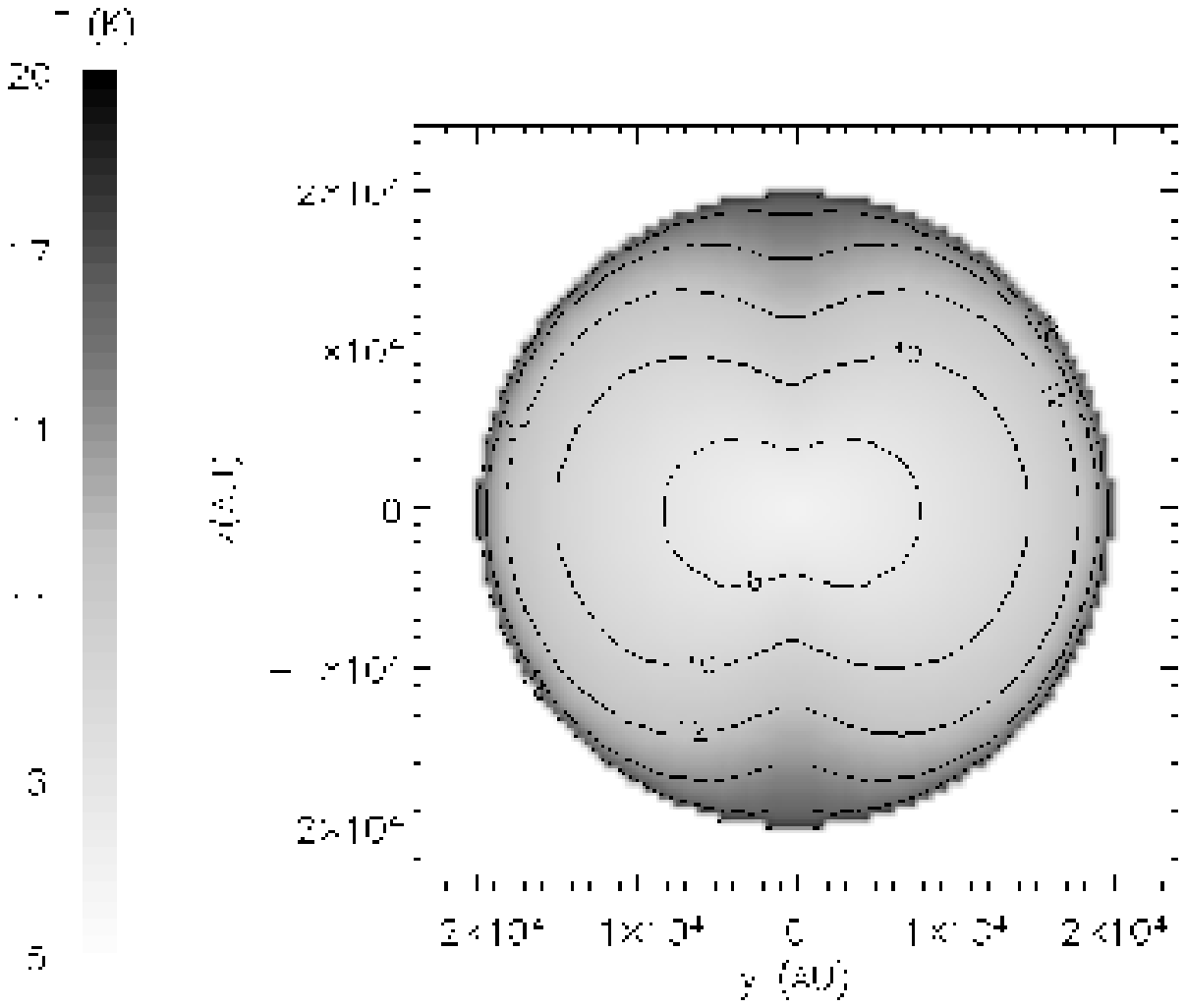}}
\vspace{-3cm}\hspace{1cm}{d}\vspace{+3cm}
\caption{Temperature distribution on the $x=0$ plane, for the models 
presented in Fig.~\ref{fig_dens.asyma.1.5}. We plot iso-temperature 
contours from 8 to 18~K, every 2~K.}
\label{fig_temp.asyma}
\label{fig_temp.asyma.1.5}
\label{fig_temp.asyma.2.5}
\label{fig_temp.asyma2.1.5}
\label{fig_temp.asyma2.2.5}
\end{figure}

\begin{figure} 
\centerline{
\includegraphics[width=7cm]{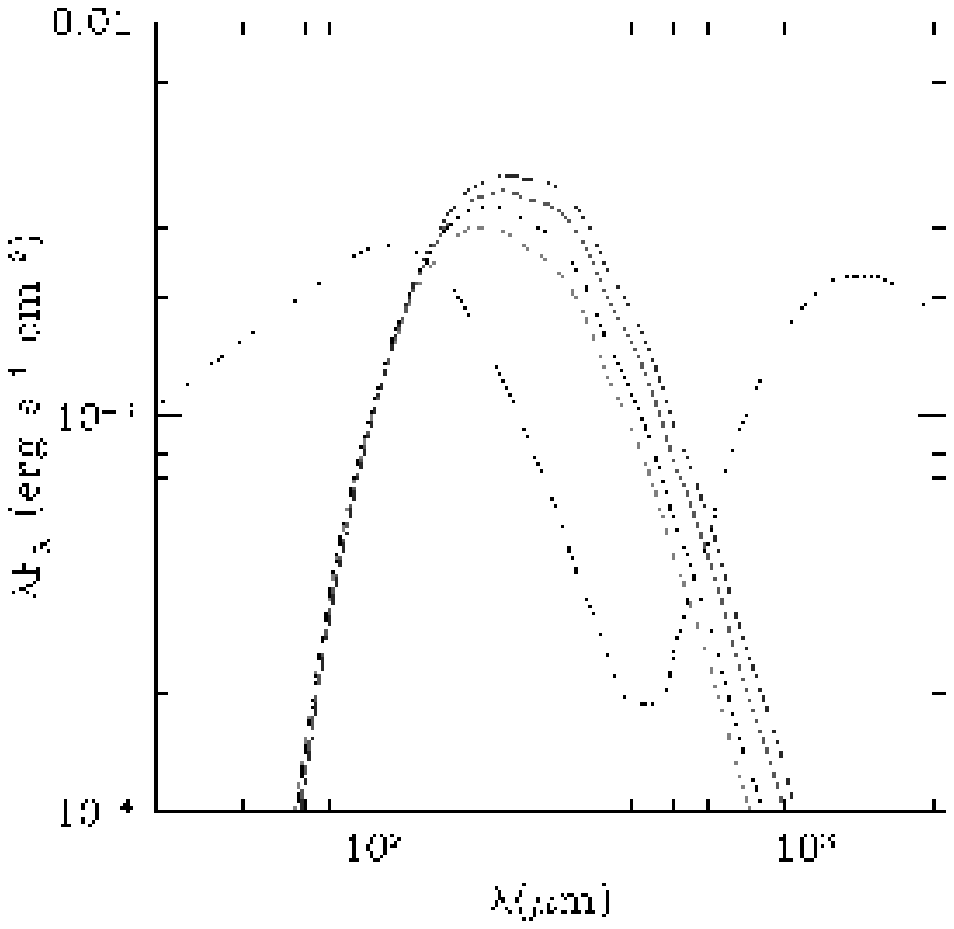}}
\caption{SED for the core models in Figs.~\ref{fig_dens.asyma},~\ref{fig_temp.asyma}; model 1.1 ($e=1.5$, 
$p=4$; short-dashed line), model 1.2 ($e=2.5$, $p=4$; solid line), 
model 1.3 ($e=1.5$, $p=1$; dash-dot line) and model 1.4 ($e=2.5$, 
$p=1$;long-dashed line). The SED of each core is independent of the 
observer's viewing angle. The dotted line on the graph corresponds to 
the incident/background SED.}
\label{fig_asyma.2.5.spec}
\end{figure}

The SED (Fig.~\ref{fig_asyma.2.5.spec}) of a specific core, for the 
model parameters we examine, is the same at any viewing angle, because 
the core is optically thin to the radiation it emits (FIR and longer 
wavelengths). Thus, it is not possible to distinguish between flattened 
and spherical cores, using SED observations, unless the core is 
extremely flattened, so that it is optically thick on the equator at 
FIR and longer wavelengths. Using the Ossenkopf \& Henning (1994) 
opacities, the minimum optical depth through the model core at 
$200\,\mu m$ is $2 \tau_{\rm V}(\theta = 0\degr) \times (\kappa_{200 \mu m}
/\kappa_{\rm V}) \simeq 0.04$, so we would need to treat much larger 
values of $e$ ($\ga 25$) in order for the SED to be 
significantly dependent on viewing angle.

In contrast, the isophotal maps of a core do depend on the observer's 
viewing angle. Additionally, they depend on the wavelength of observation. 
Our code calculates images at any wavelength, and therefore provides a 
useful tool for direct comparison with observations, e.g.  at mid-infrared 
(ISO/ISOCAM), far-infrared (ISO/ISOPHOT) and submm/mm (SCUBA, IRAM) 
wavelengths. We distinguish two wavelength regions on which we focus: 
(i) wavelengths near the peak of the core emission (150-250$~\micron$; 
we choose 200$~\micron$ as a representative wavelength), and (ii)
wavelengths much longer than the peak (submm amd mm region; we choose
850~$\micron$ as a representative wavelength). In each of the above 
regions the isophotal maps have similar general characteristics.

At 200~$\micron$ the core appearance depends both on its temperature and its 
column density in the observer's direction. It is seen in  
Figs.~\ref{images.asyma.1.5}-\ref{images.asyma2.2.5}, that cores with disk-like 
asymmetry appear spherical when viewed pole-on and flattened when viewed edge-on. 
The outer parts of a core can be more or less luminous than the central 
parts, depending on the core temperature and the observer's viewing 
angle. For example, at close to pole-on viewing angles the outer parts 
of the core are  more luminous than the inner parts of the core (limb 
brightening; e.g. Figs.~\ref{images.asyma.1.5}-\ref{images.asyma2.2.5}, 
$\theta=$0\degr). This happens because the temperature is higher in the 
outer parts and this more than compensates for the lower column density 
(since at wavelengths near the peak of the core emission and shorter, 
the Planck intensity $ B_\nu (T)$  depends on temperature as 
$B_\nu (T)\propto e^{-a/T}$, $a=$const). At other viewing angles the 
appearance of the core is determined by a combination of temperature and 
column density effects (Figs.~\ref{images.asyma.1.5}-\ref{images.asyma2.2.5}, 
$\theta$=30\degr, 60\degr, 90\degr). This interplay between core temperature  
and column density along the line of sight results in characteristic 
features on the images of the cores. Such features include (i) the two 
intensity minima  at almost symmetric positions relative to the centre of 
the core, on the images at 30\degr  
(Fig.~\ref{images.asyma.2.5}-\ref{images.asyma2.2.5}), and (ii) the two 
intensity maxima, again at  symmetric positions relative to the centre 
of the core, on the images at 90\degr 
(Figs.~\ref{images.asyma.1.5}-\ref{images.asyma.2.5}). (It is also 
worth mentioning that although the characteristic features appear in 
symmetric positions relative to both axes of density-symmetry, we 
should expect deviations from symmetry to arise if the radiation field  
incident on the core is not isotropic.)

\begin{figure}
\centerline{
\includegraphics[width=7cm]{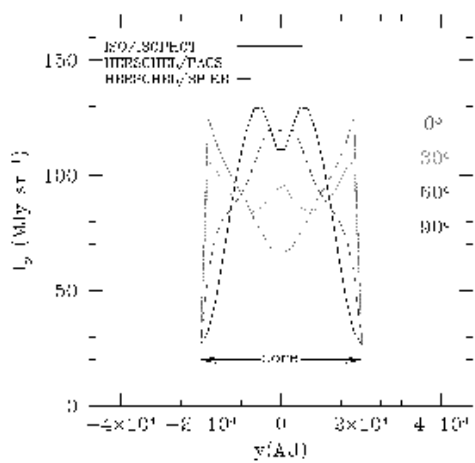}}
\caption{A perpendicular cut through the centre of the core images 
presented in Fig.~\ref{images.asyma.2.5} for model 1.2 at 200~$\micron$ 
(but also including the background radiation field). In addition we 
plot the beam size of ISO/ISOPHOT (90\arcsec $\equiv$ 9000~AU for a 
core at 100 pc) and the beam size of the upcoming (2007) {\it Herschel}
(13\arcsec $\equiv$ 1300~AU for the $170~\micron$ band of PACS and 
17\arcsec $\equiv$ 1700~AU for the $250~\micron$ band of SPIRE). ISO's 
resolution may not be good enough to detect any of the features on the 
graph, but {\it Herschel} will have a  much better resolution and should 
be able to detect such features in the future.} 
\label{image_rprof_nem}
\end{figure}

\begin{figure}
\centerline{\includegraphics[width=5.5cm,angle=-90]{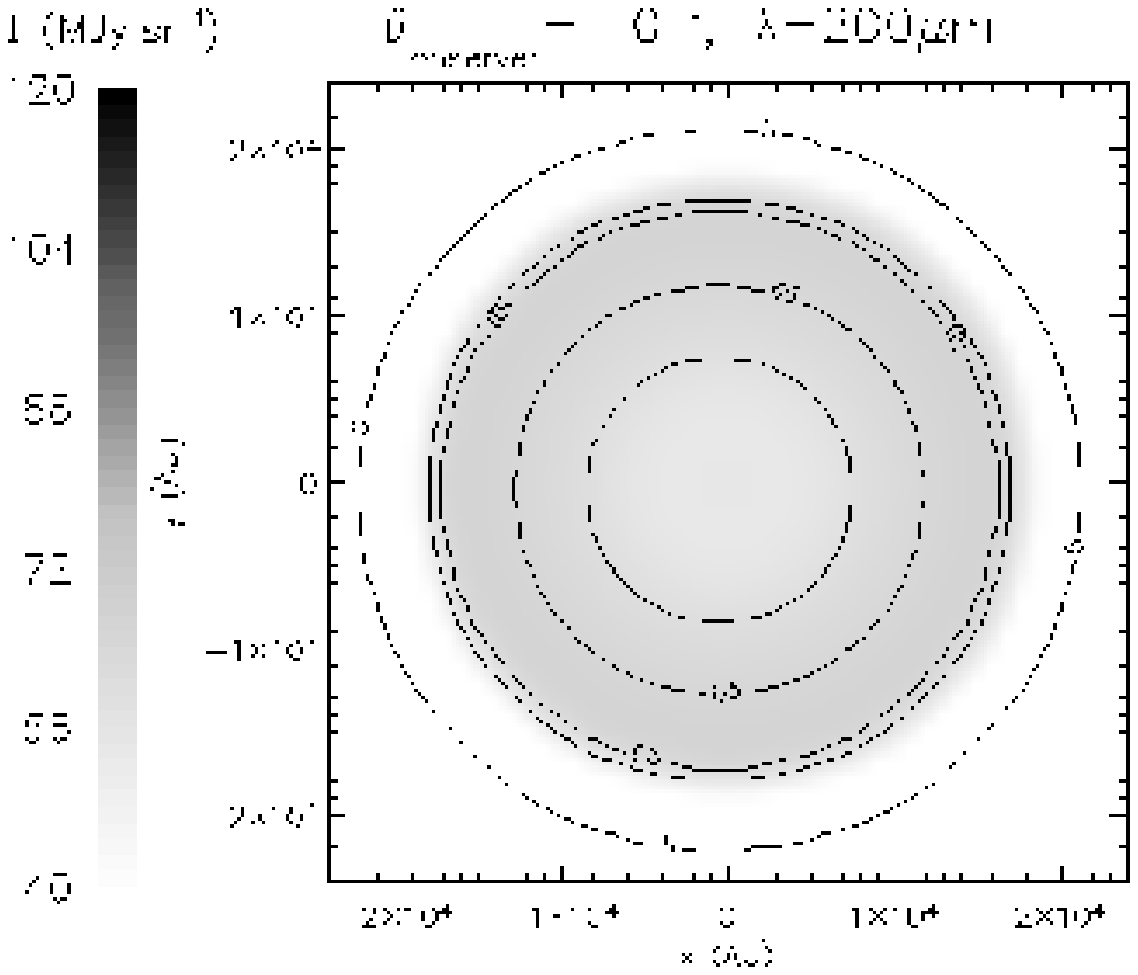}}
\centerline{\includegraphics[width=5.7cm,angle=-90]{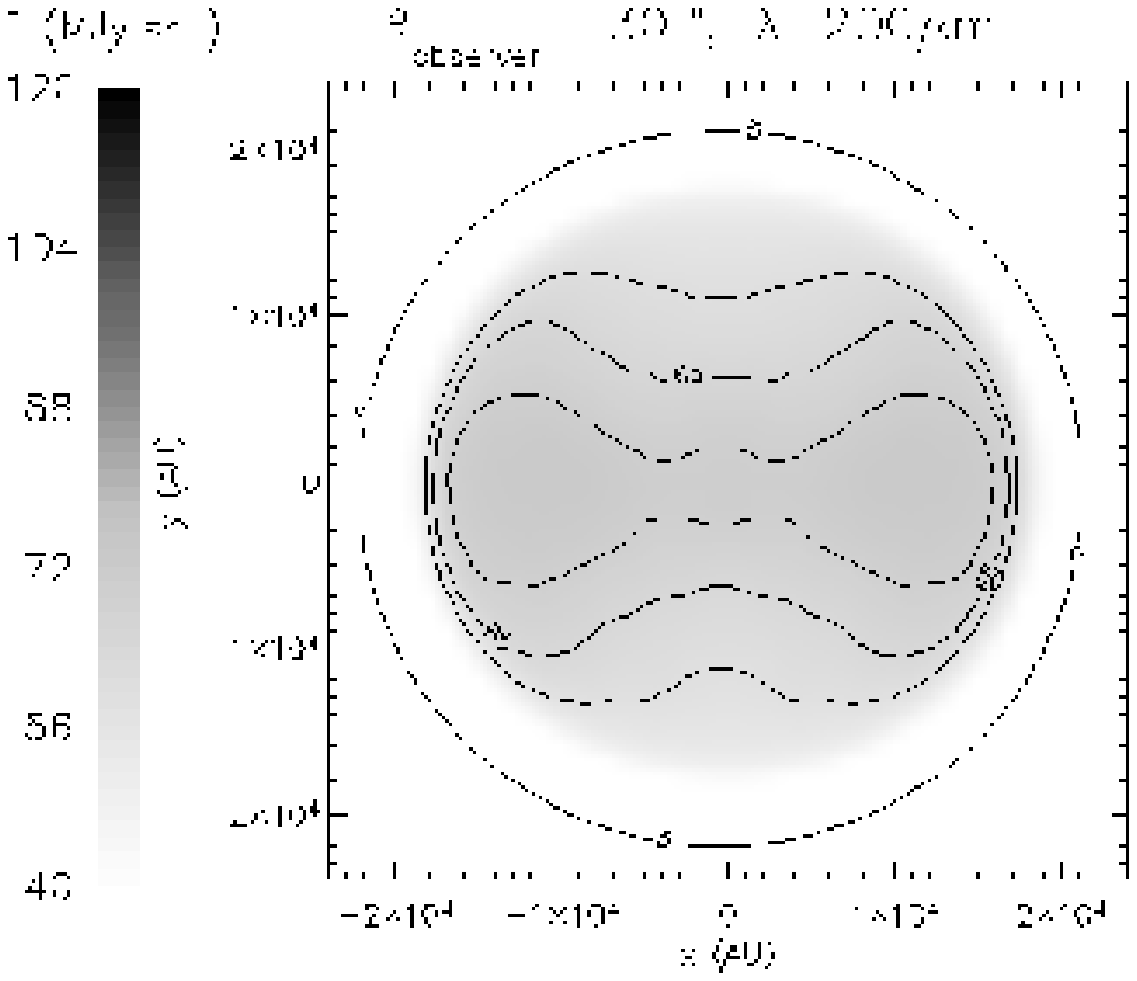}}
\centerline{\includegraphics[width=5.7cm,angle=-90]{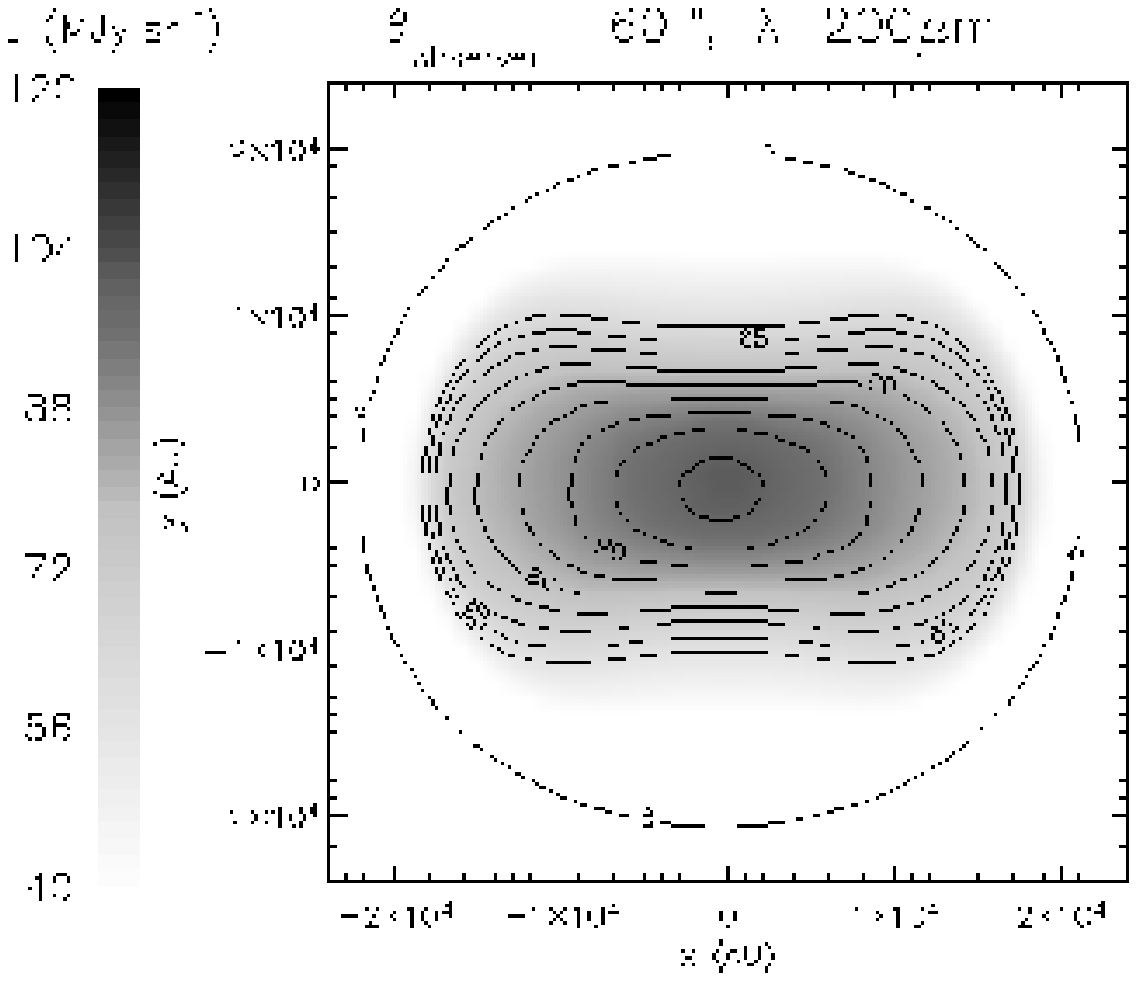}}
\centerline{\includegraphics[width=5.7cm,angle=-90]{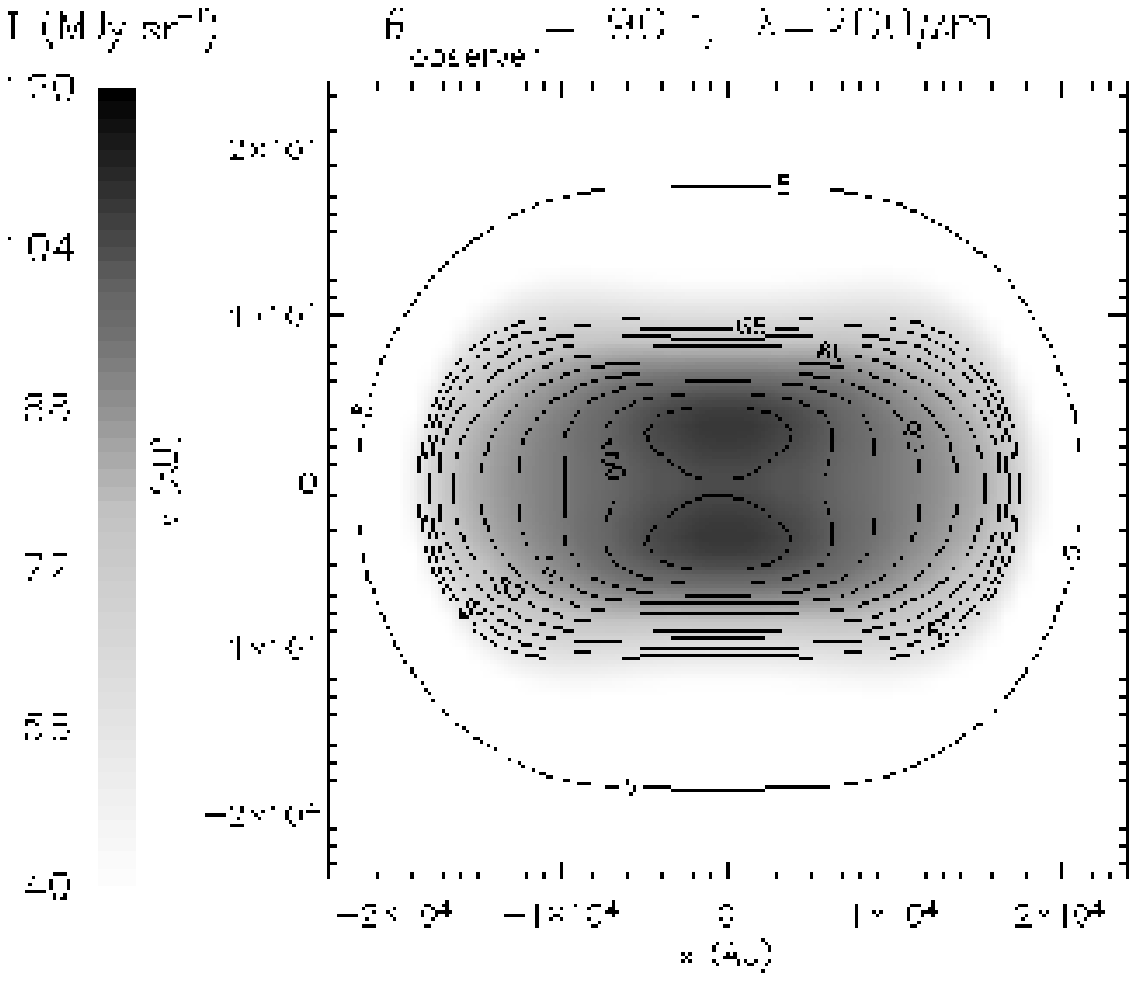}}
\caption{Isophotal maps at 200~$\micron$ at viewing angles $0\degr$, 
$30\degr$, $60\degr$ and $90\degr$, for a flattened core with 
equatorial-to-polar optical depth ratio $e=1.5$  and $p=4$ (model 1.1). 
We plot an isophotal contour at 5~MJy~sr$^{-1}$ and then from 60 to
110~MJy~sr$^{-1}$, every 5~MJy~sr$^{-1}$. There are characteristic 
symmetric features due to core temperature and orientation with respect 
to the observer. We note that in these isophotal maps, and in all 
subsequent isophotal maps, the axes $(x,y)$ refer to the plane of sky 
as seen by the observer.} 
\label{images.asyma.1.5}
\end{figure}

\begin{figure}
\centerline{\includegraphics[width=5.7cm,angle=-90]{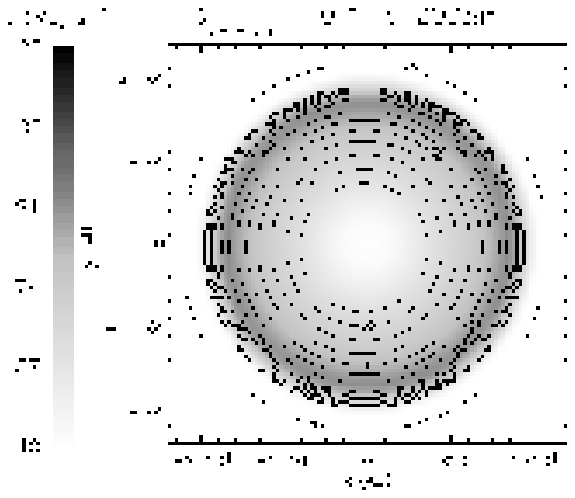}}
\centerline{\includegraphics[width=5.7cm,angle=-90]{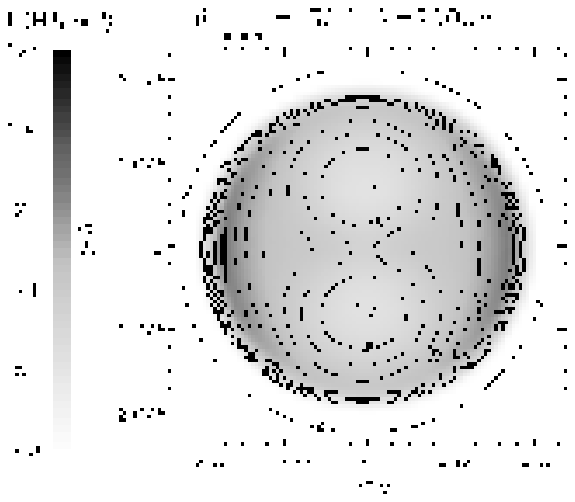}}
\centerline{\includegraphics[width=5.7cm,angle=-90]{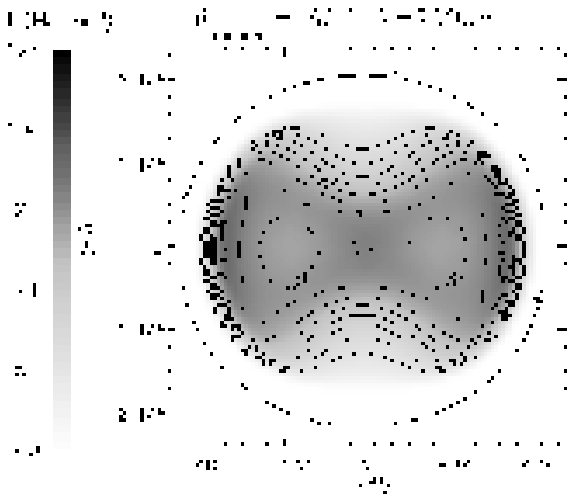}}
\centerline{\includegraphics[width=5.7cm,angle=-90]{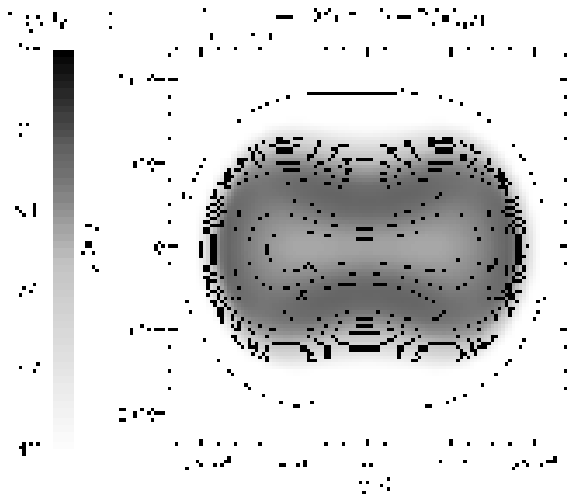}}
\caption{Same as  Fig.~\ref{images.asyma.1.5}, but for a more flattened 
core, with equatorial-to-polar optical depth ratio  $e=2.5$ and $p=4$ 
(model 1.2).} 
\label{images.asyma.2.5}
\end{figure}

\begin{figure}
\centerline{\includegraphics[width=5.7cm,angle=-90]{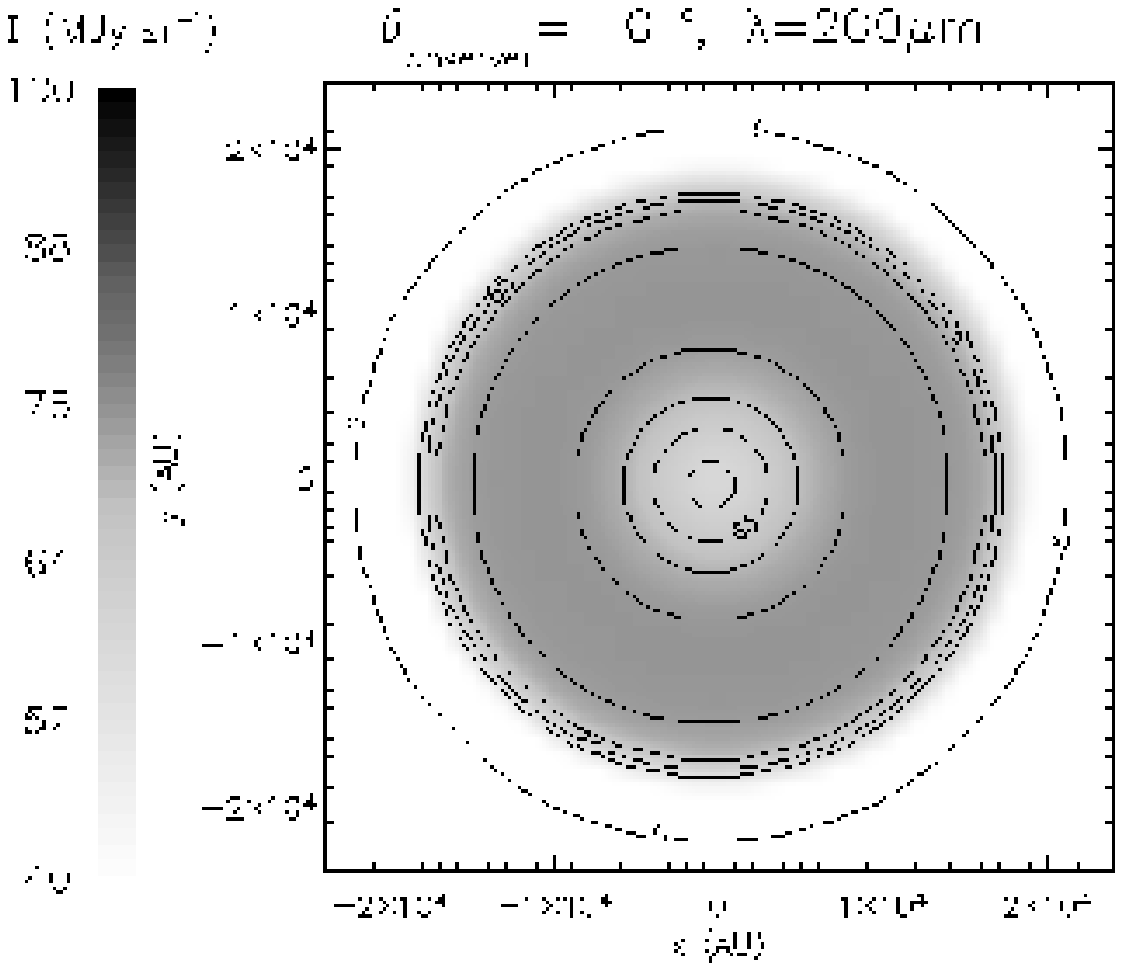}}
\centerline{\includegraphics[width=5.7cm,angle=-90]{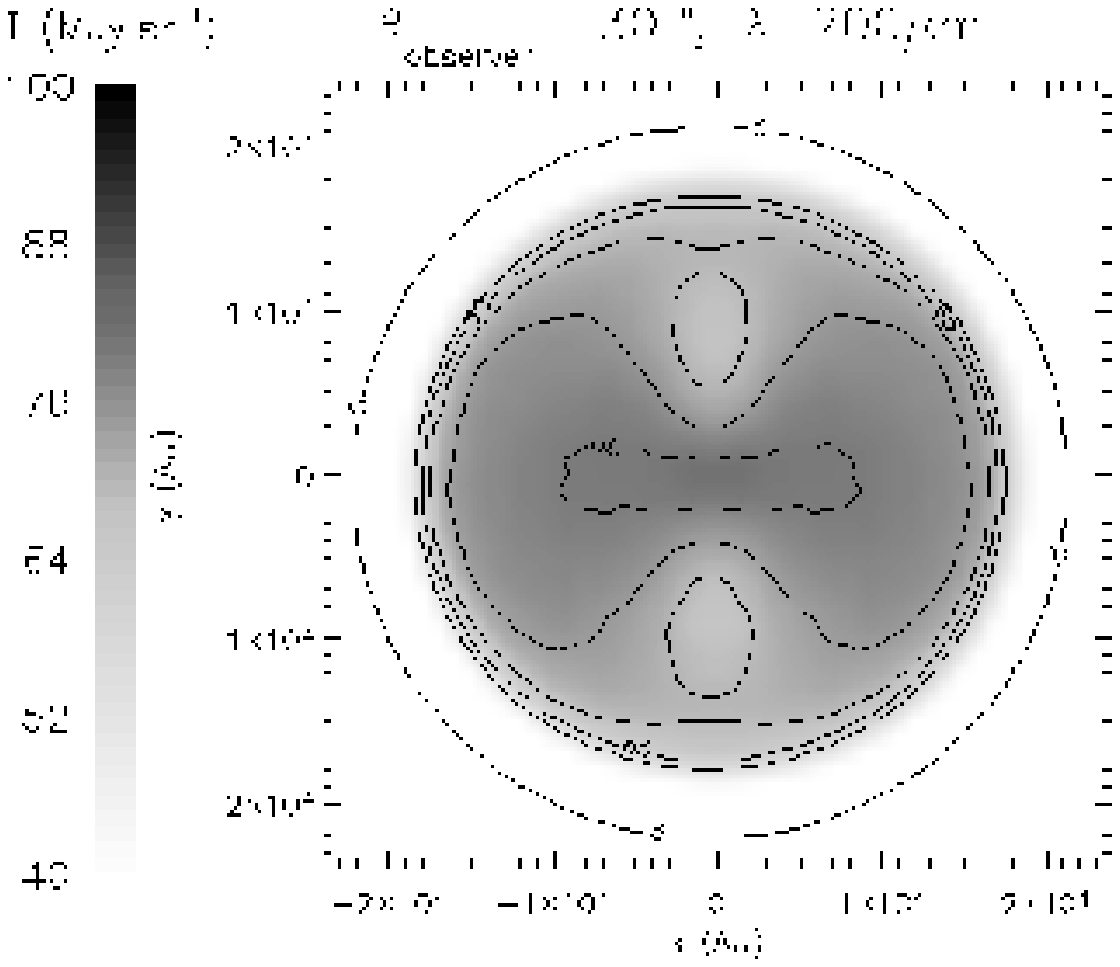}}
\centerline{\includegraphics[width=5.7cm,angle=-90]{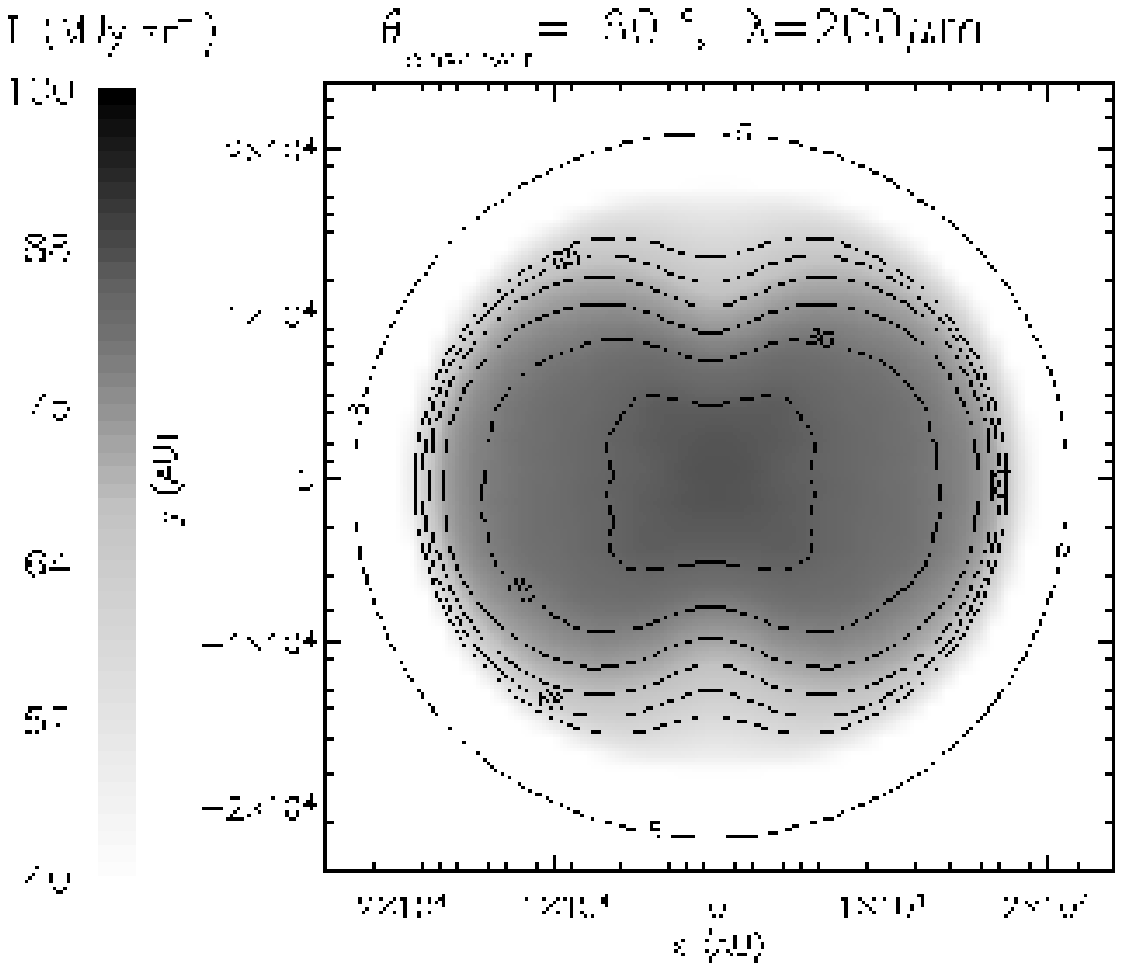}}
\centerline{\includegraphics[width=5.7cm,angle=-90]{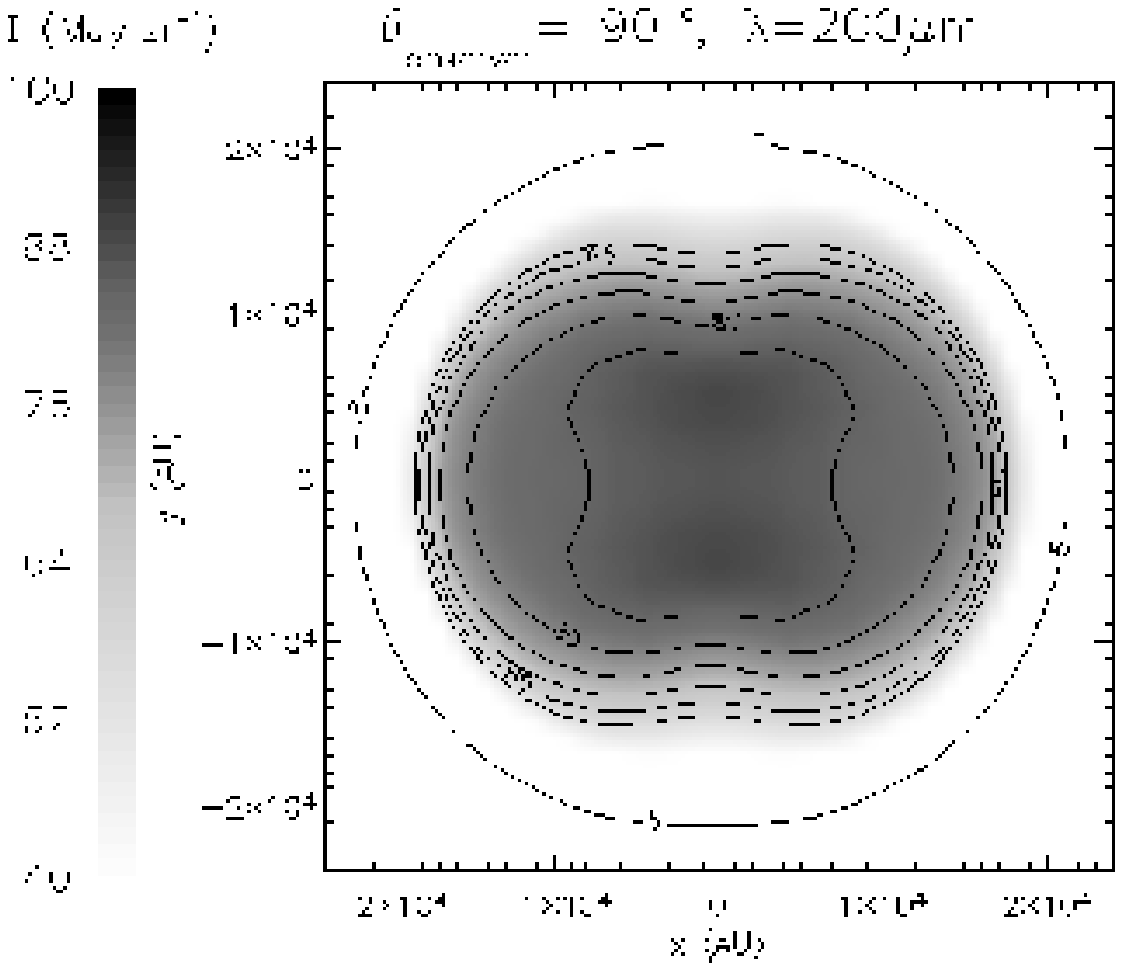}}
\caption{Same as  Fig.~\ref{images.asyma.1.5}, but for a  core with 
equatorial-to-polar optical depth ratio  $e=1.5$ and $p=1$  (model 1.3).} 
\label{images.asyma2.1.5}
\end{figure}

\begin{figure}
\centerline{\includegraphics[width=5.7cm,angle=-90]{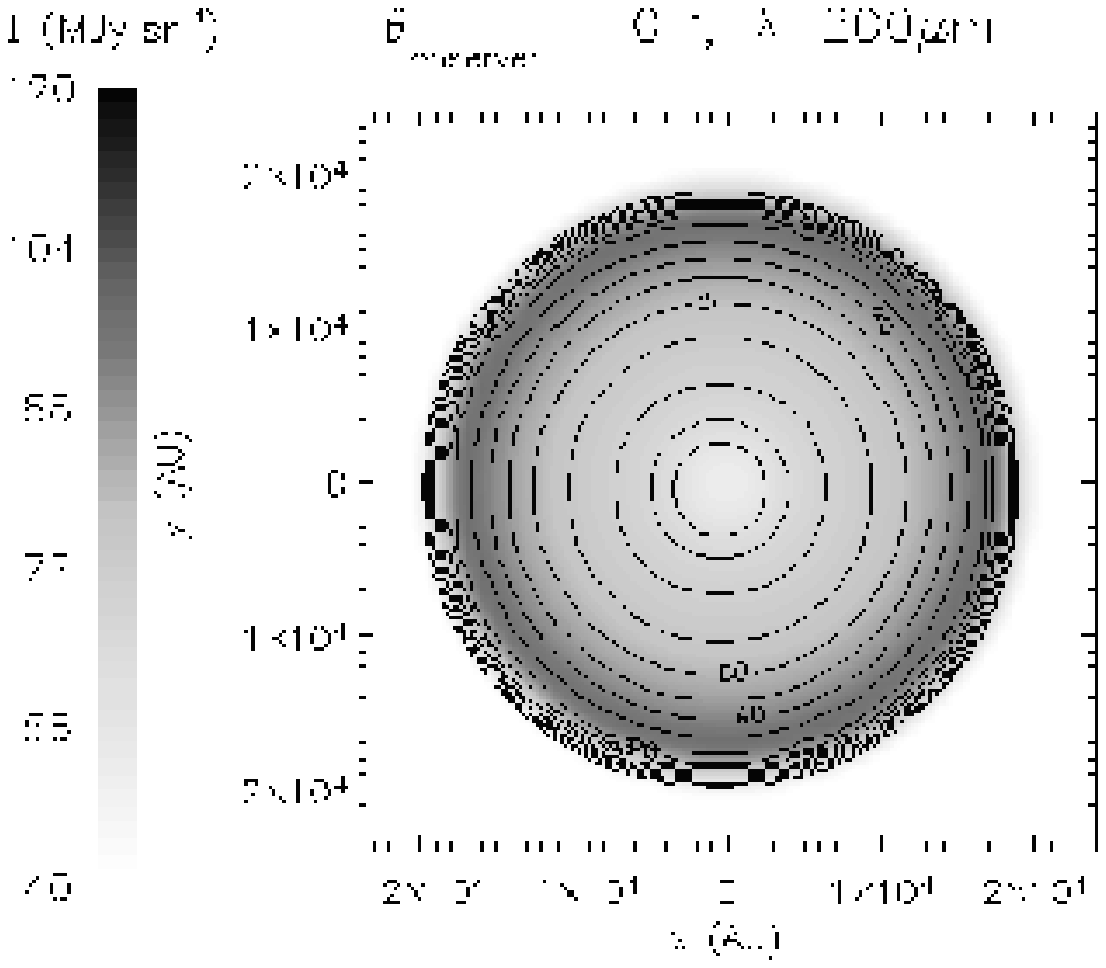}}
\centerline{\includegraphics[width=5.7cm,angle=-90]{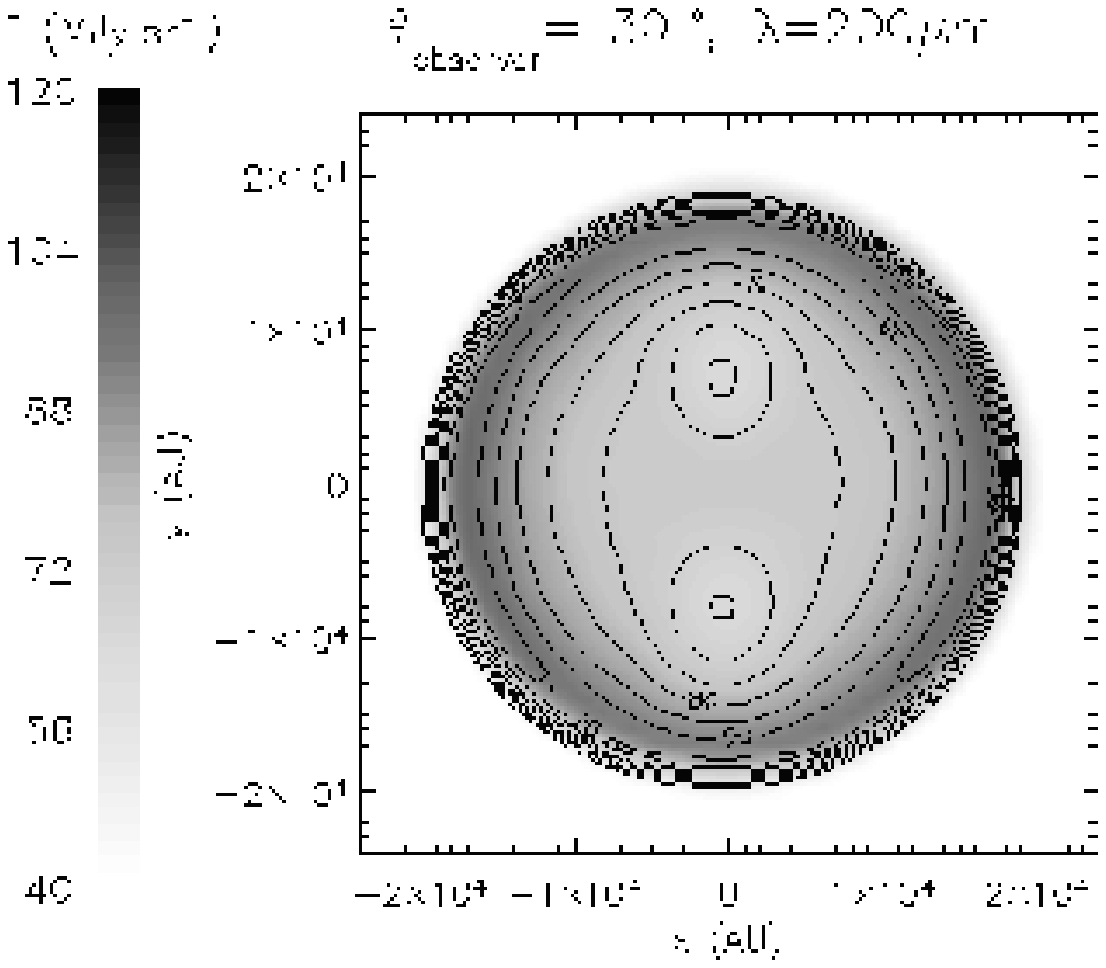}}
\centerline{\includegraphics[width=5.7cm,angle=-90]{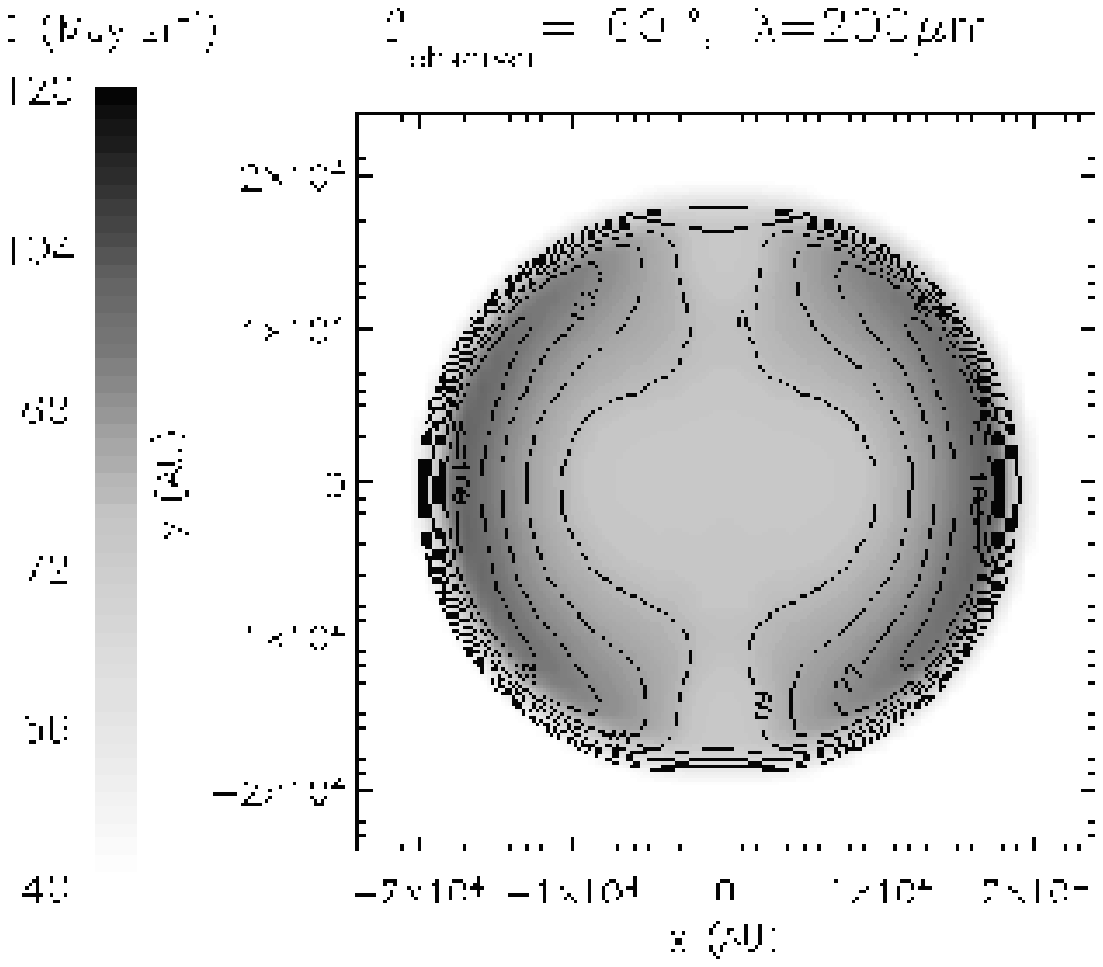}}
\centerline{\includegraphics[width=5.7cm,angle=-90]{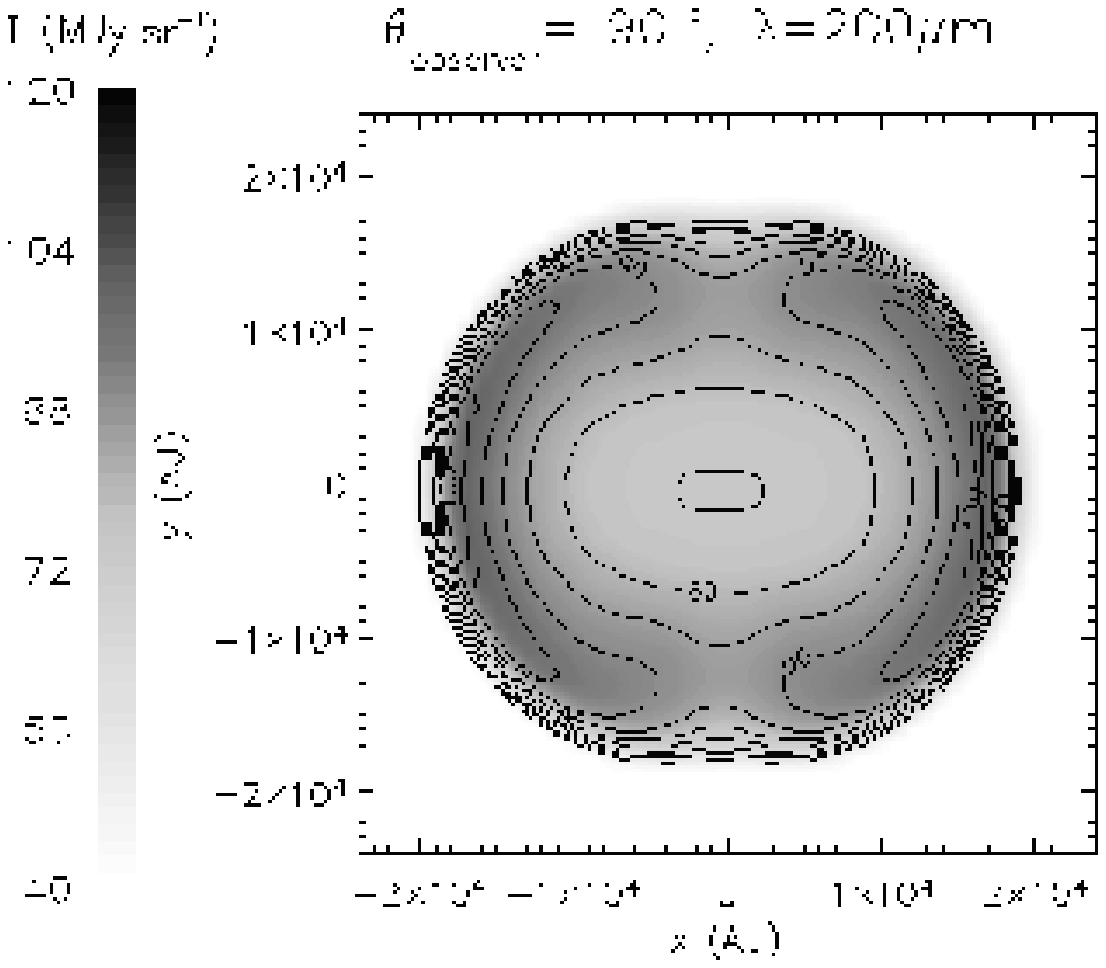}}
\caption{Same as Fig.~\ref{images.asyma.1.5}, but for a  core with 
equatorial-to-polar optical depth ratio  $e=2.5$ and $p=1$  (model 1.4).}
\label{images.asyma2.2.5}
\end{figure}

We conclude that  isophotal maps at 200~$\micron$ contain detailed  
information, and sensitive, high resolution observations at 
200~$\micron$, could be helpful in constraining the core density 
and temperature structure and  the orientation of the core with 
respect to the observer. In Fig.~\ref{image_rprof_nem}, we present 
a perpendicular cut through the centre of the core images shown in 
Fig.~\ref{images.asyma.2.5}. We also plot the beam size of the 
ISOPHOT C-200 camera (90\arcsec, or 9000~AU for a core at 100 pc) 
and the beam size of the upcoming (2007) {\it Herschel} (13\arcsec 
or 1300~AU for the $170~\micron$ band of PACS; 17\arcsec or 1700~AU 
for the $250~\micron$ band of SPIRE). ISOPHOT's resolution is probably 
not good enough to detect the features mentioned above. Indeed, a 
search in the Kirk et al. (2004) sample of ISO/ISOPHOT observations 
(also see Ward-Thompson et. al 2002) does not reveal any cores with 
such distinctive features. However, {\it Herschel} should, in 
principle, be able to detect such features in the future.

\begin{figure}
\centerline{\includegraphics[width=5.7cm,angle=-90]{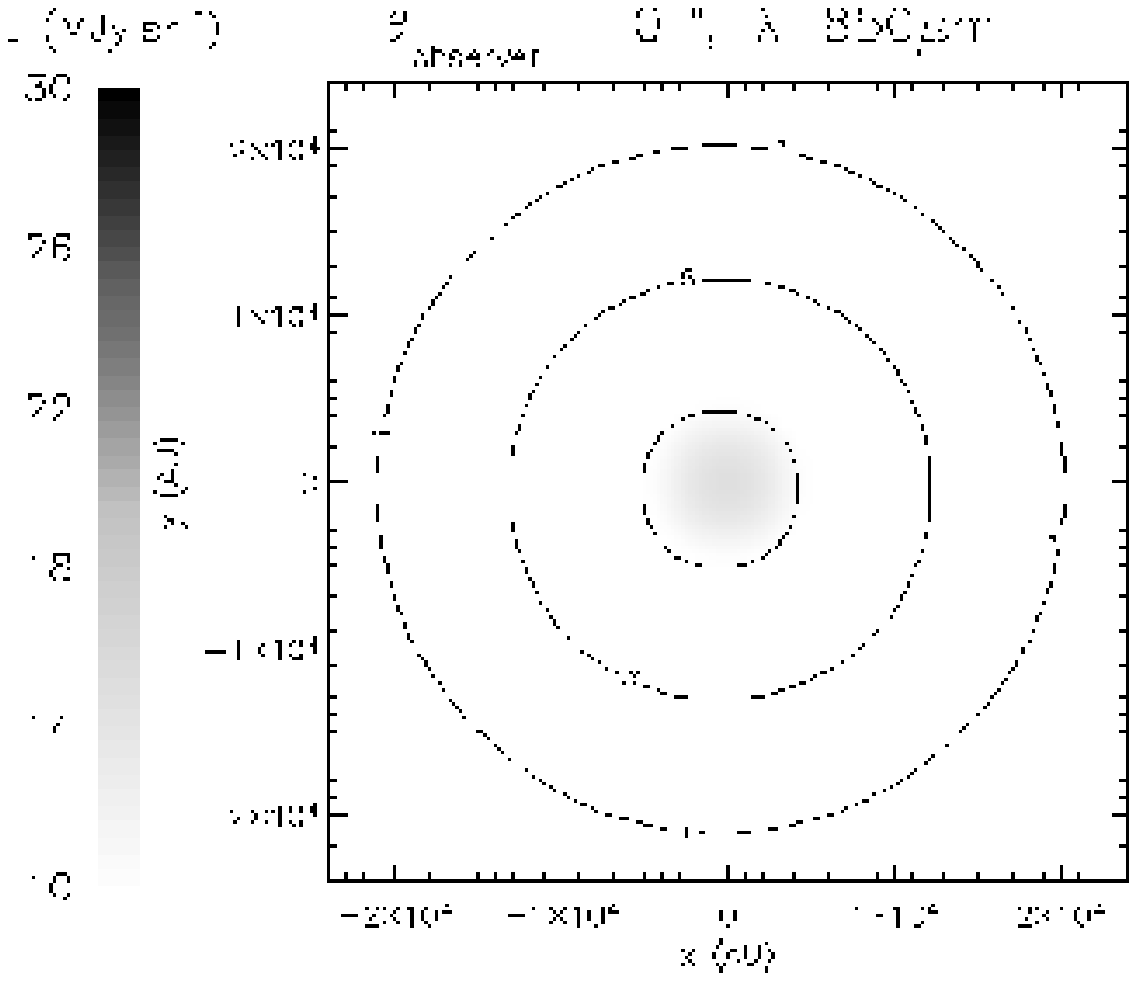}}
\centerline{\includegraphics[width=5.7cm,angle=-90]{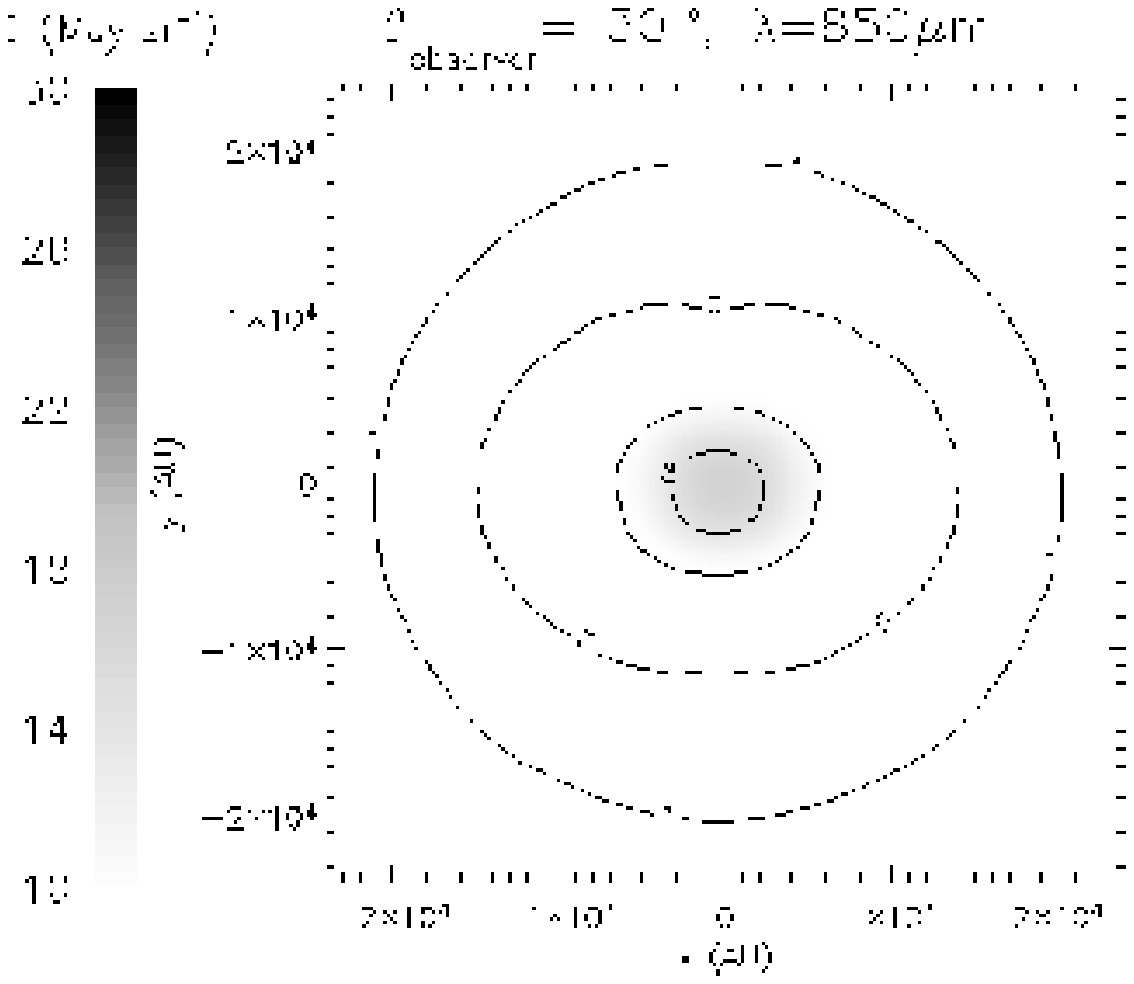}}
\centerline{\includegraphics[width=5.7cm,angle=-90]{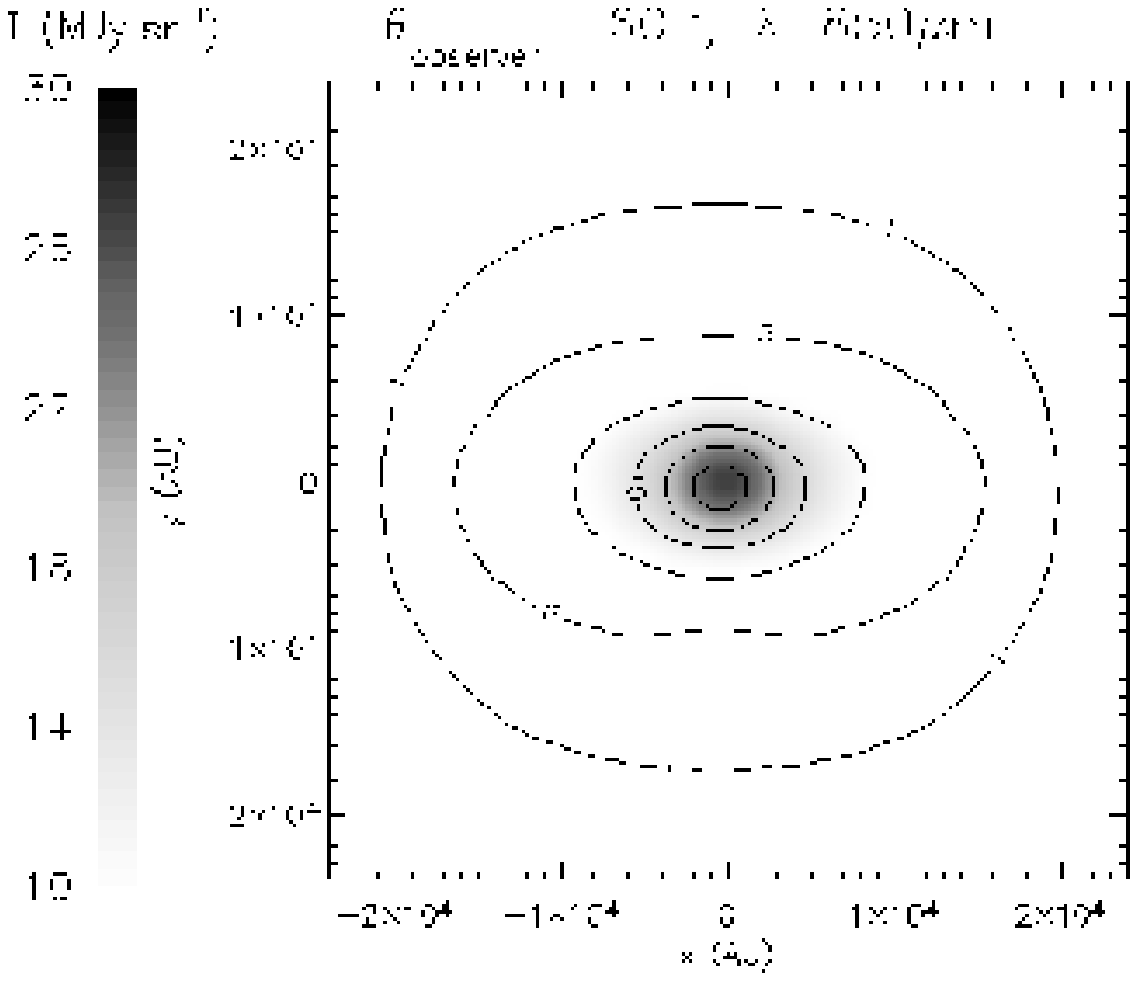}}
\centerline{\includegraphics[width=5.7cm,angle=-90]{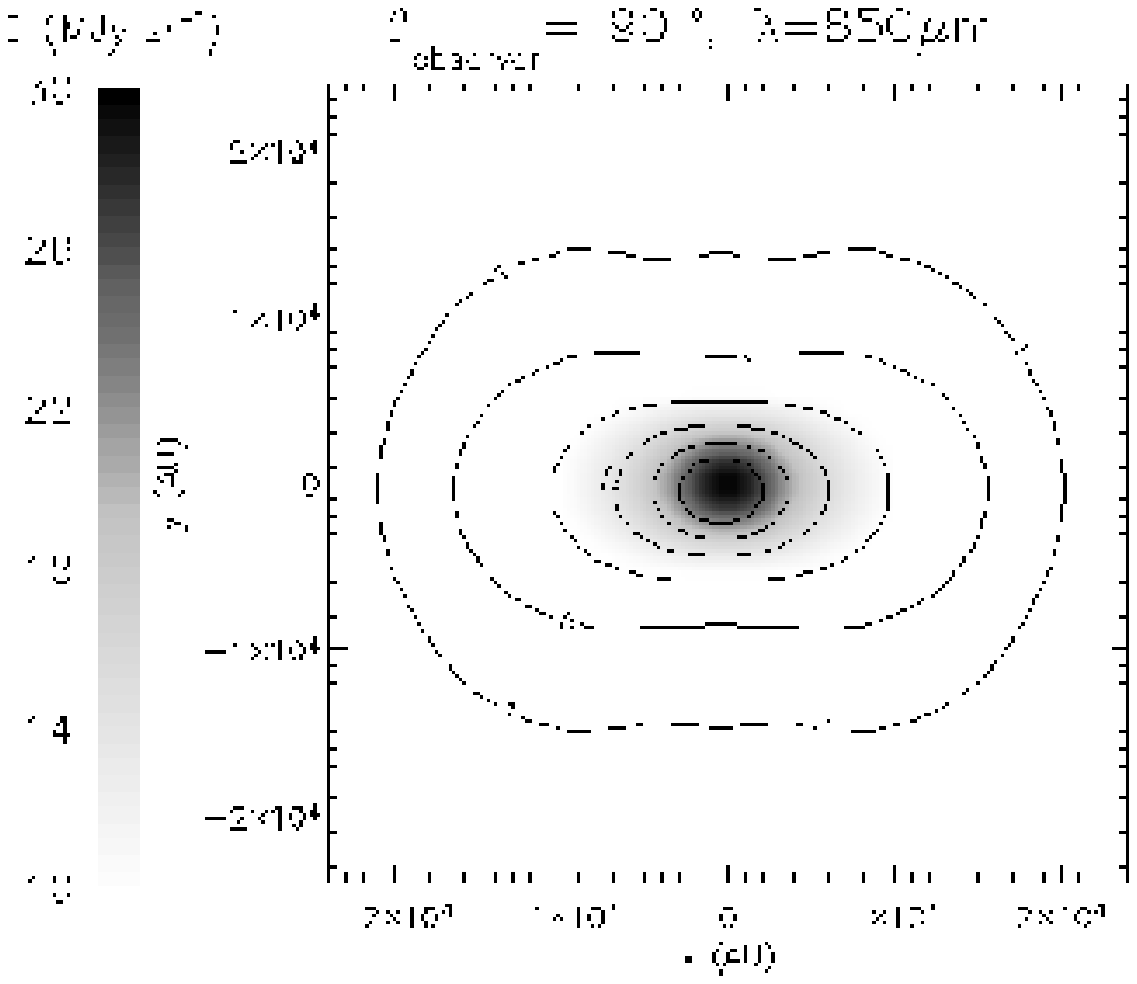}}
\caption{Isophotal maps at 850~$\micron$ at viewing angles $0\degr$, 
$30\degr$, $60\degr$ and $90\degr$, for a flattened core with 
equatorial-to-polar  optical depth ratio $e=1.5$ and $p=4$ (model 
1.1). We plot an isophotal contour at 1~MJy~sr$^{-1}$ and then from 
5 to 50~MJy~sr$^{-1}$, every 5~MJy~sr$^{-1}$. The core appears 
elongated when viewed at an angle other than $\theta=0\degr$.}
\label{images.asyma.1.5.850}
\end{figure}

\begin{figure}
\centerline{\includegraphics[width=5.7cm,angle=-90]{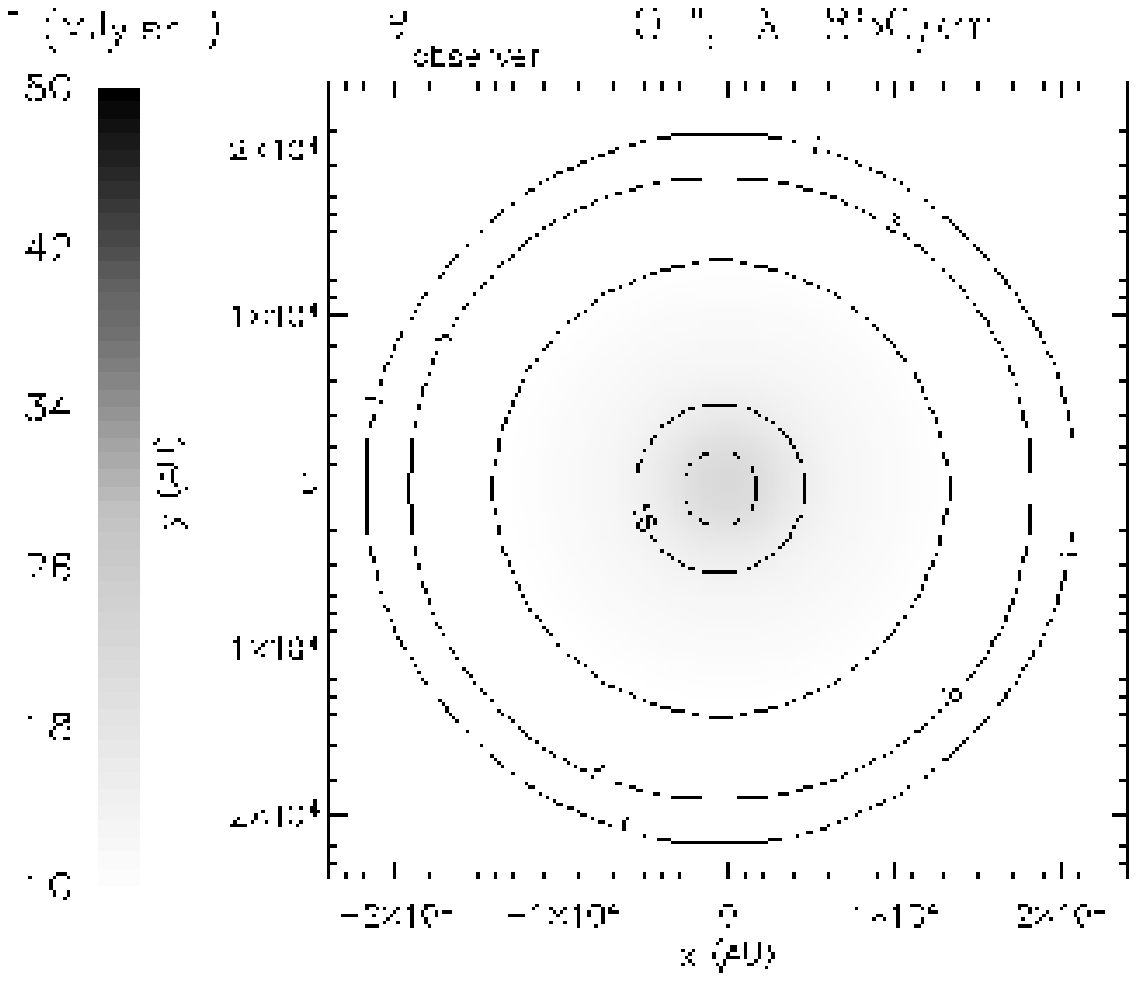}}
\centerline{\includegraphics[width=5.7cm,angle=-90]{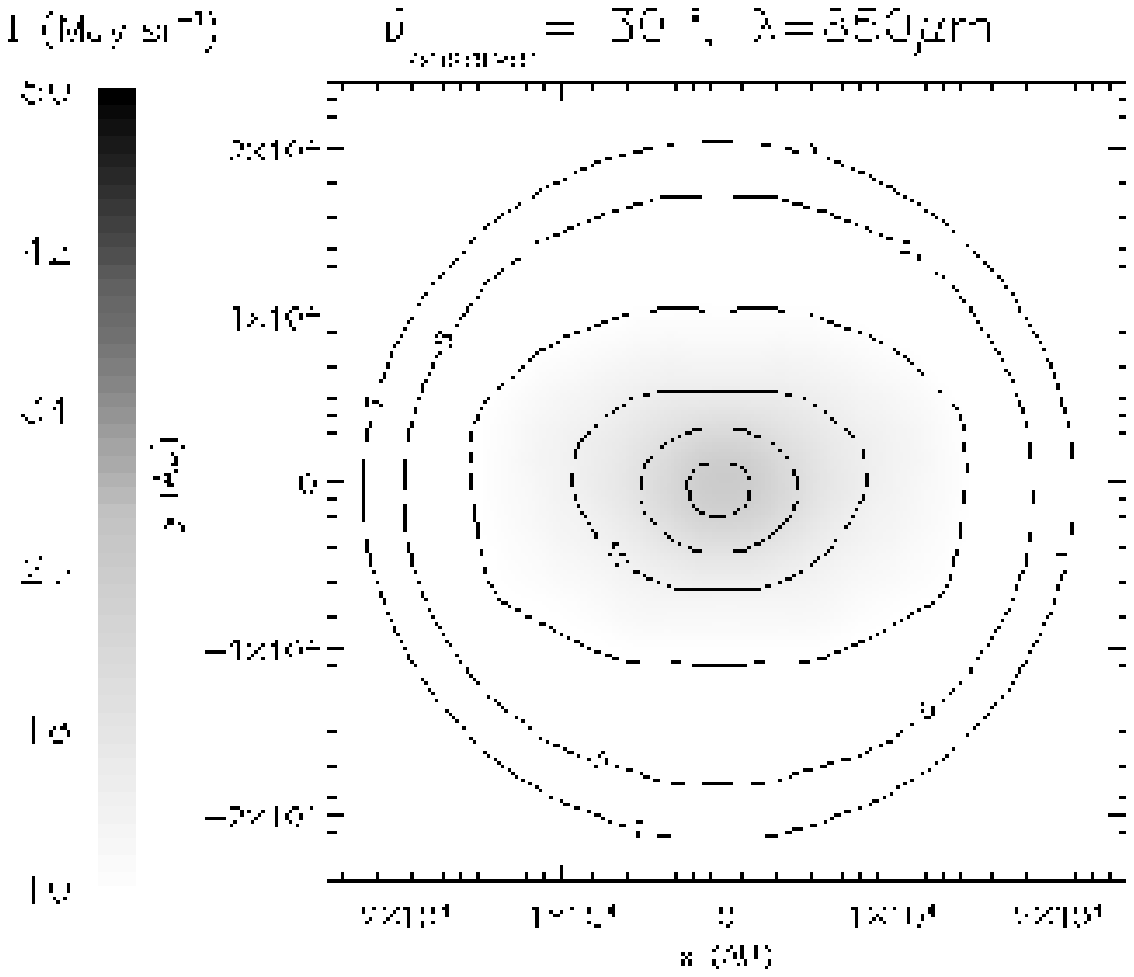}}
\centerline{\includegraphics[width=5.7cm,angle=-90]{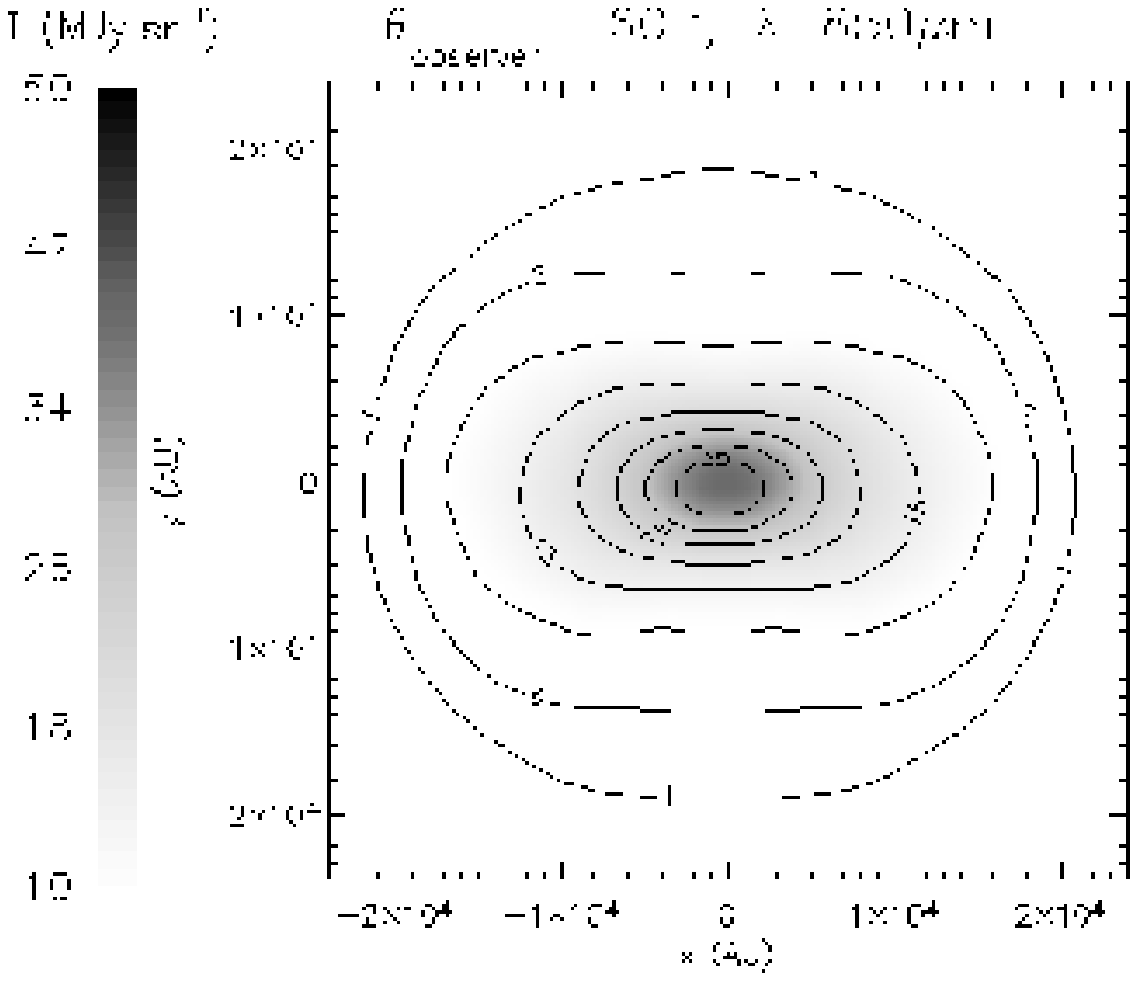}}
\centerline{\includegraphics[width=5.7cm,angle=-90]{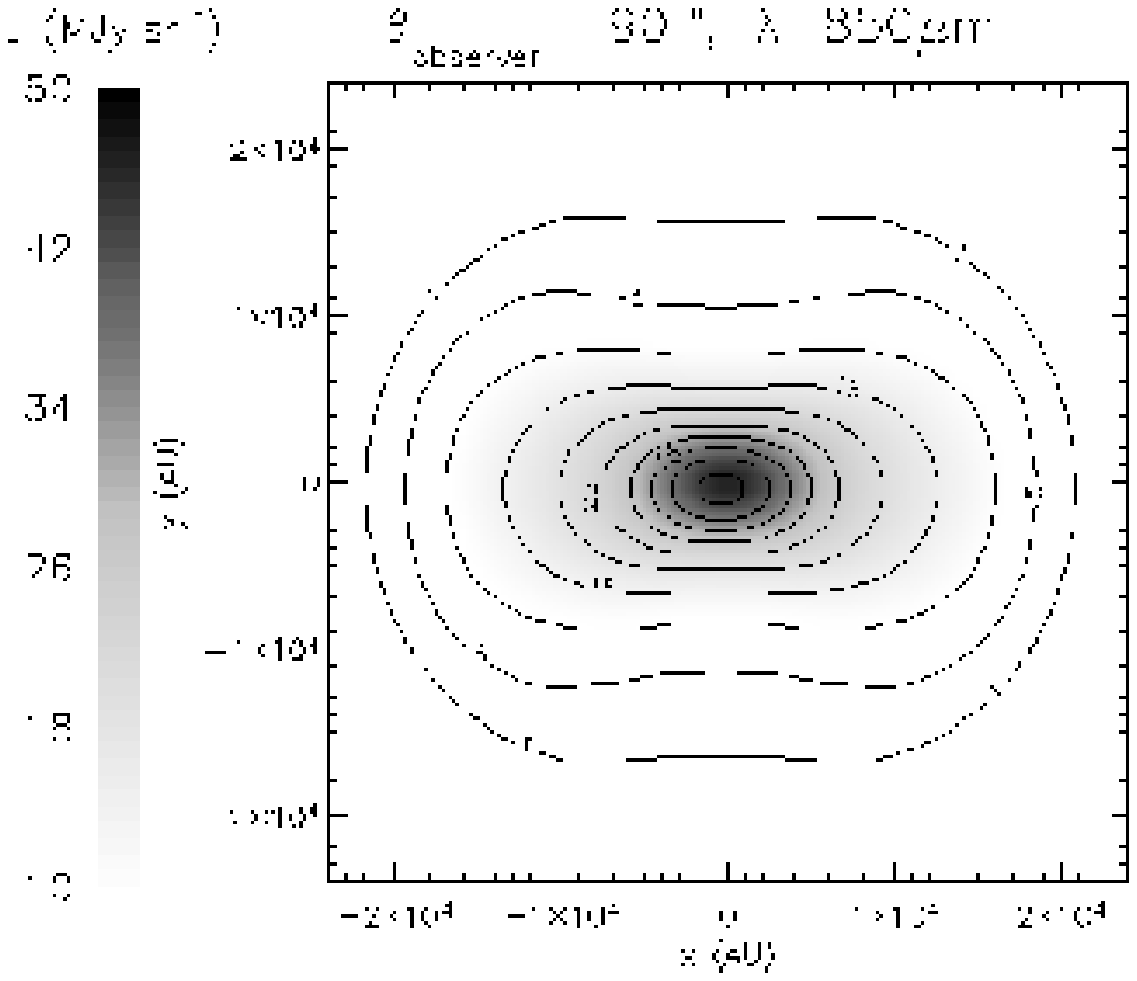}}
\caption{Same as Fig.~\ref{images.asyma.1.5.850}, but for a more flattened 
core, with equatorial-to-polar optical depth ratio  $e=2.5$ and $p=4$ 
(model 1.2).}
\label{images.asyma.2.5.850} 
\end{figure}

\begin{figure}
\centerline{\includegraphics[width=5.7cm,angle=-90]{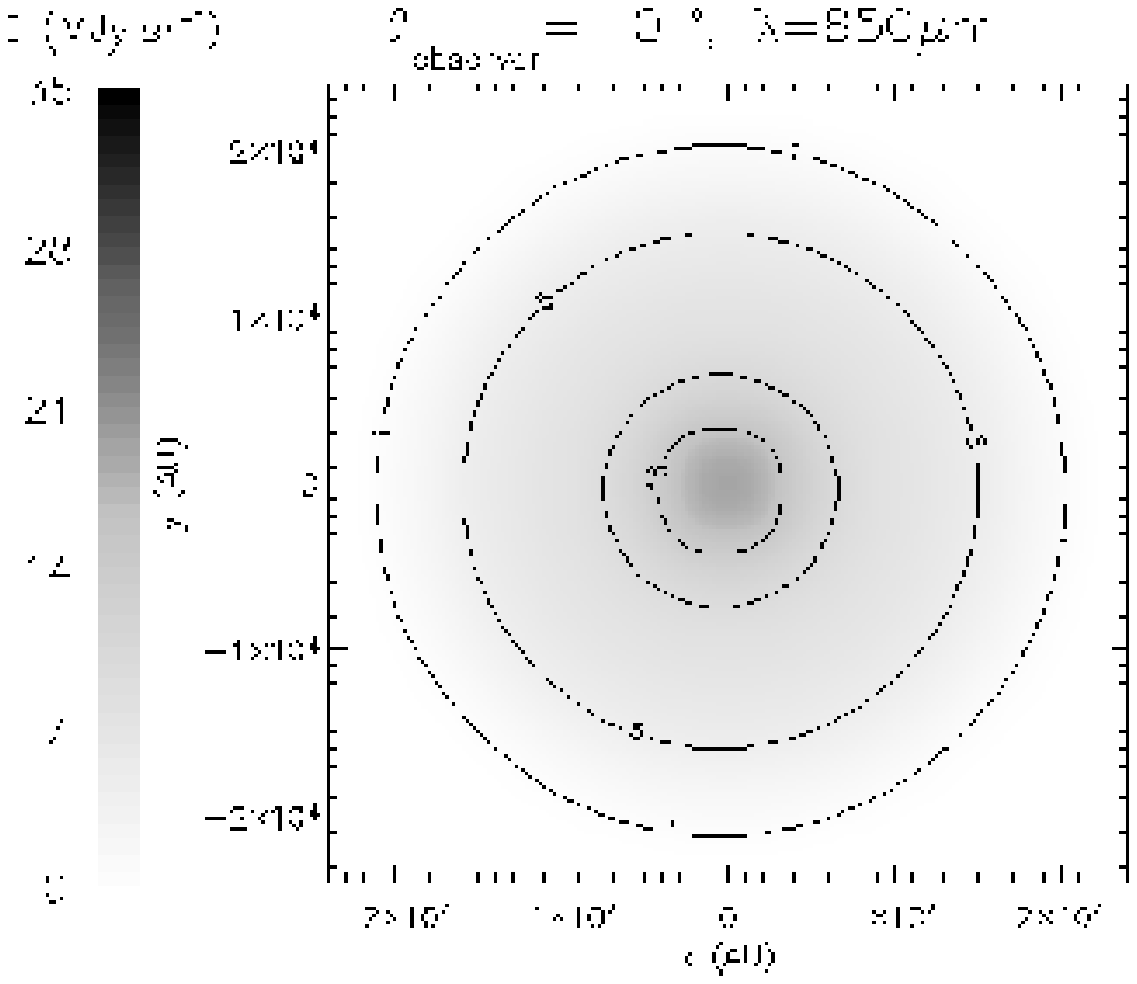}}
\centerline{\includegraphics[width=5.7cm,angle=-90]{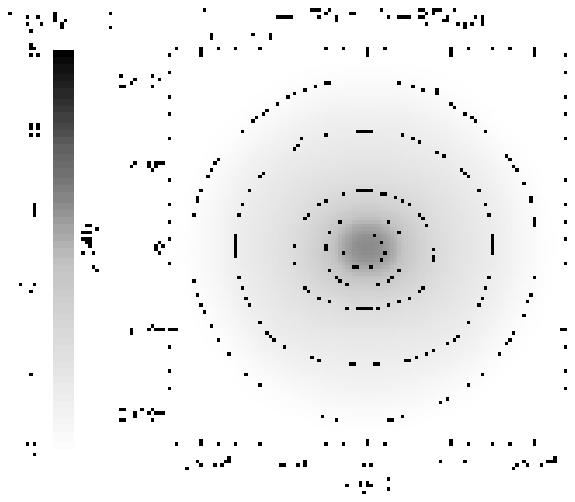}}
\centerline{\includegraphics[width=5.7cm,angle=-90]{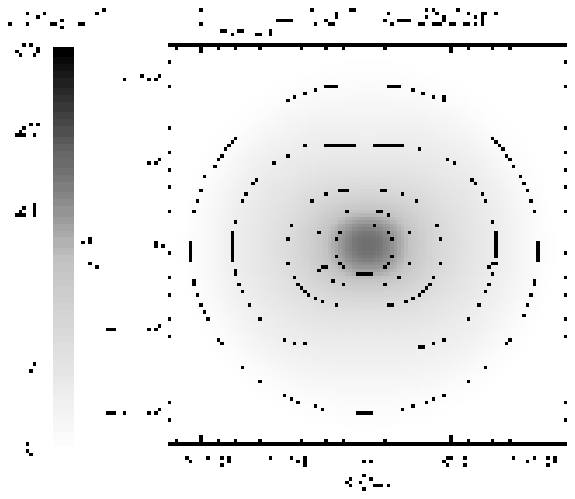}}
\centerline{\includegraphics[width=5.7cm,angle=-90]{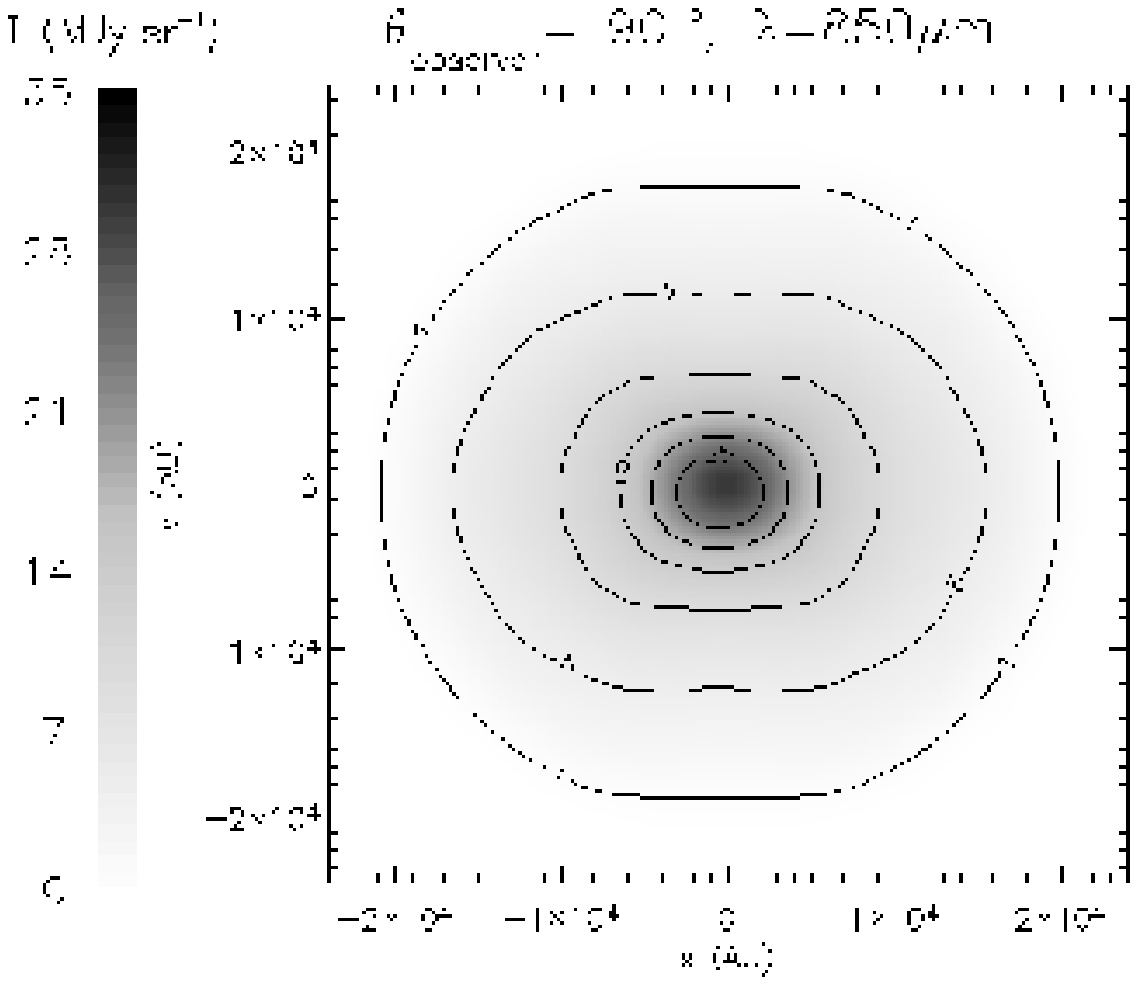}}
\caption{Same as  Fig.~\ref{images.asyma.1.5.850}, but for a  core with 
equatorial-to-polar optical depth ratio  $e=1.5$ and $p=1$  (model 1.3).}
\label{images.asyma2.1.5.850}
\end{figure}

\begin{figure}
\centerline{\includegraphics[width=5.7cm,angle=-90]{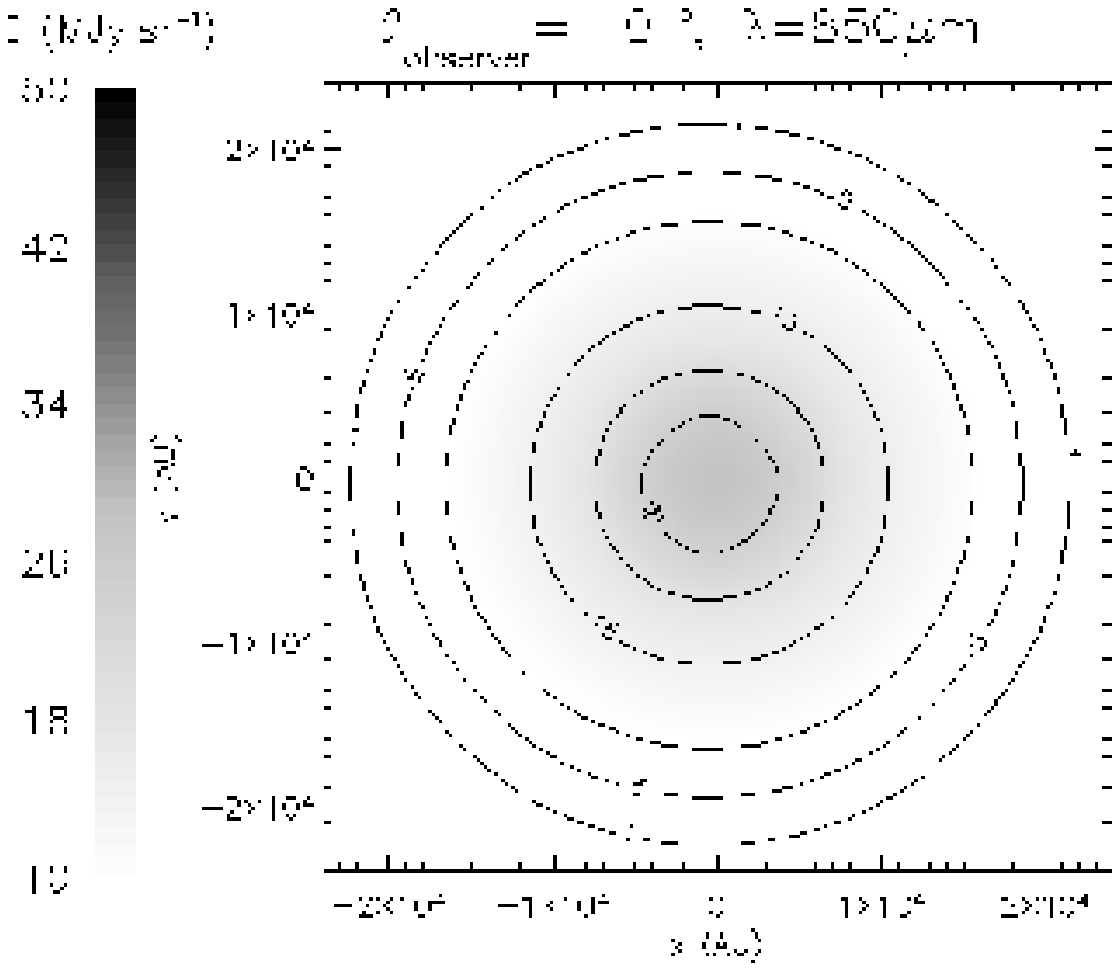}}
\centerline{\includegraphics[width=5.7cm,angle=-90]{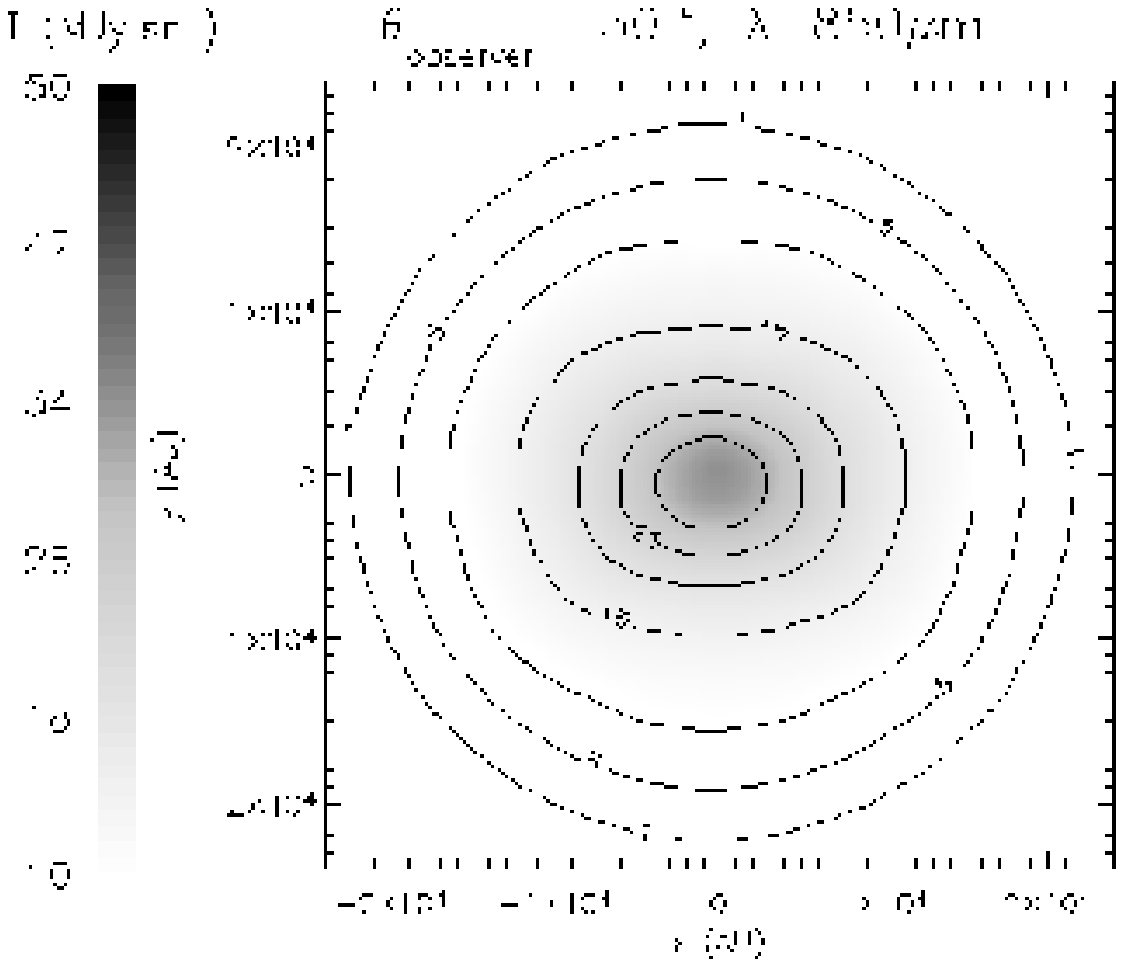}}
\centerline{\includegraphics[width=5.7cm,angle=-90]{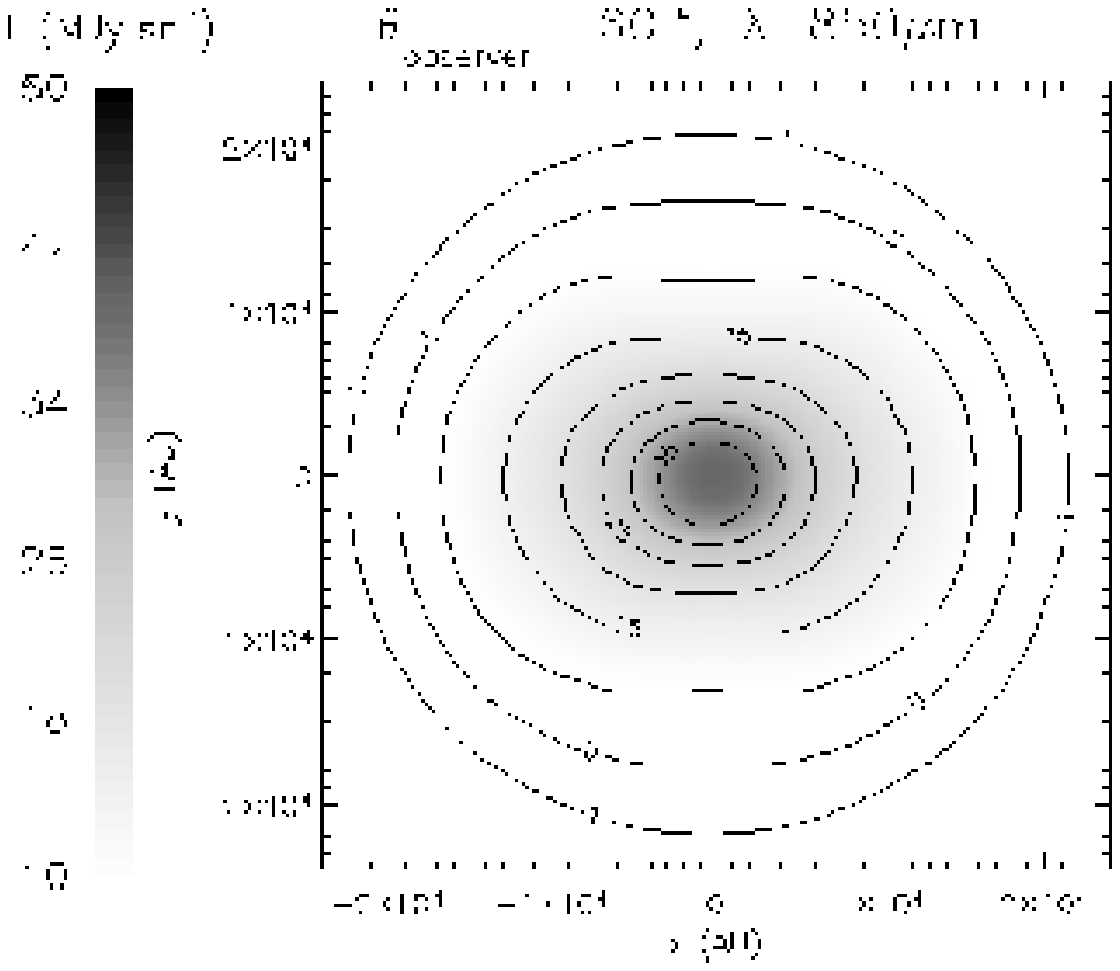}}
\centerline{\includegraphics[width=5.7cm,angle=-90]{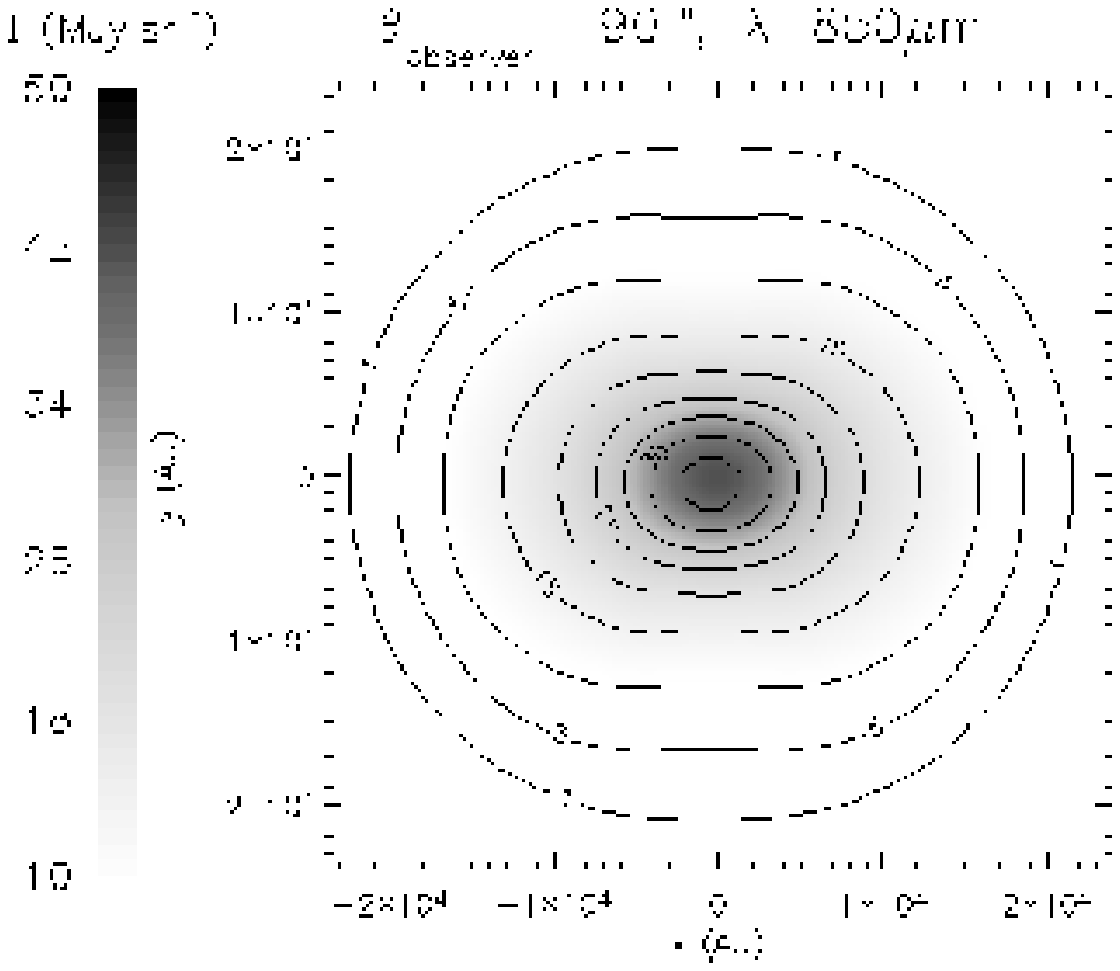}}
\caption{Same as  Fig.~\ref{images.asyma.1.5.850}, but for a  core with 
equatorial-to-polar optical depth ratio  $e=2.5$ and $p=1$  (model 1.4).}
\label{images.asyma2.2.5.850}
\end{figure}

In the second wavelength region (submm and mm wavelengths) the core 
emission is mainly regulated by the column density (e.g. at 850~$\micron$, 
Figs.~\ref{images.asyma.1.5.850}-\ref{images.asyma2.2.5.850}). Thus, the 
intensity is larger at the centre, where the column density is larger. 
For the same reason the core appears flattened when the observer looks 
at it from any direction other than pole-on. It is also evident that the 
peak intensity of the core is much larger when the core is viewed edge-on. 
Therefore, flattened cores are more prominent when viewed edge-on. This 
introduces a possible observational selection effect which should be 
taken into account when studying the shape  statistics of prestellar 
cores. Low-mass flattened cores are more likely to be detected if they 
are edge-on, than if they are face-on. This is true for optically thin 
mm and submm continuum observations, and also for optically thin molecular 
line observations. For example, when comparing the projected shapes of 
condensations from  hydrodynamic simulations with the observed shapes 
using solely optically thin continuum or optically thin molecular line 
observations, one should expect a lower number of observed near-spherical 
cores than indicated by the simulations. This may be the reason for the 
small excess of high axis ratio cores in the simulations by Gammie et al. 
(2003) (see their Fig.~9).

\subsection{Embedded prestellar cores}

In the previous section, we studied cores that are  directly exposed to 
the interstellar radiation field (as  approximated by the BISRF). However, 
cores are generally embedded in molecular clouds, with visual optical 
depths ranging from 2-10 (e.g. in Taurus) up to 40 (e.g. in $\rho$ 
Ophiuchi). The ambient  cloud absorbs the energetic UV and optical 
photons and re-emits them in the FIR and submm (because the ambient cloud 
is generally cold, $T_{\rm cloud}\sim20-100$~K). Therefore, the radiation 
incident on a core that is embedded in a cloud is reduced in the UV and 
optical, and enhanced in  the FIR and submm (Mathis et al. 1983). Previous 
radiative transfer calculations of spherical cores embedded at the centre 
of an ambient cloud (Stamatellos \& Whitworth 2003a), have shown that 
embedded cores are colder ($T<12$~K) and that the temperature gradients 
inside these cores are smaller than in non-embedded cores. Andr\'e et al. 
(2003) also found that the temperatures inside embedded cores are lower 
than in non-embedded cores (assuming that they are heated by the same 
ISRF), using a different approach, in which they estimated the effective 
radiation field incident on an embedded core from observations.

\begin{figure}[h]
\centerline{
\includegraphics[width=6cm]{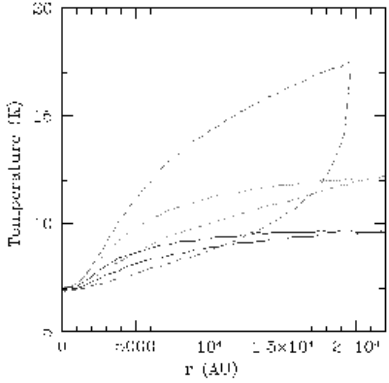}}
\caption{The effect of the parent cloud on cores. Temperature profiles 
of a non-embedded core (model 1.2; dashed lines), and of a core at the 
centre of an ambient cloud with visual extinction $A_{\rm V}=4$ (dotted 
lines), and $A_{\rm V}=13$ (solid lines). The upper curve of each set of 
lines corresponds to the direction towards the pole of the core 
($\theta=0\degr$), and the bottom curve to the direction towards the 
core equator ($\theta=90\degr$). The core is colder when it resides 
inside a thicker parent cloud (i.e. when it is illuminated by a 
radiation field that is  weakened at short ($<10\micron$) wavelengths), 
and the temperature differences between different parts of the core 
are smaller. Thus, the characteristic features on the isophotal maps 
at wavelengths near the peak of the core SED are weakened.}
\label{fig_asym_a.temp.comp}
\end{figure}

\begin{figure}[h]
\centerline{
\includegraphics[width=6cm,angle=-90]{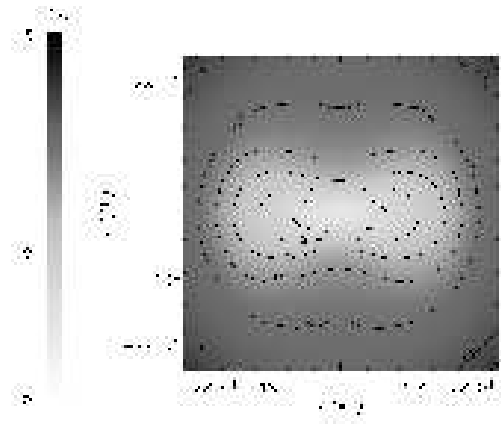}}
\caption{Temperature distribution on the $x=0$ plane, for the same model 
presented in Figs.~\ref{fig_dens.asyma.2.5}b and \ref{fig_temp.asyma.2.5}b 
($e=2.5$, $p=4$, model 1.2), but embedded in the centre of an ambient 
molecular cloud with visual extinction $A_{\rm V}=4$. We plot 
iso-temperature contours from 8 to 13~K, every 1~K. The embedded core is 
colder than the non-embedded core (Fig.~\ref{fig_temp.asyma.2.5}b) and 
the temperature gradient inside the core is smaller.}
\label{fig_temp.asyma.2.5.em}
\end{figure}

\begin{figure}[h]
\centerline{
\includegraphics[width=7cm]{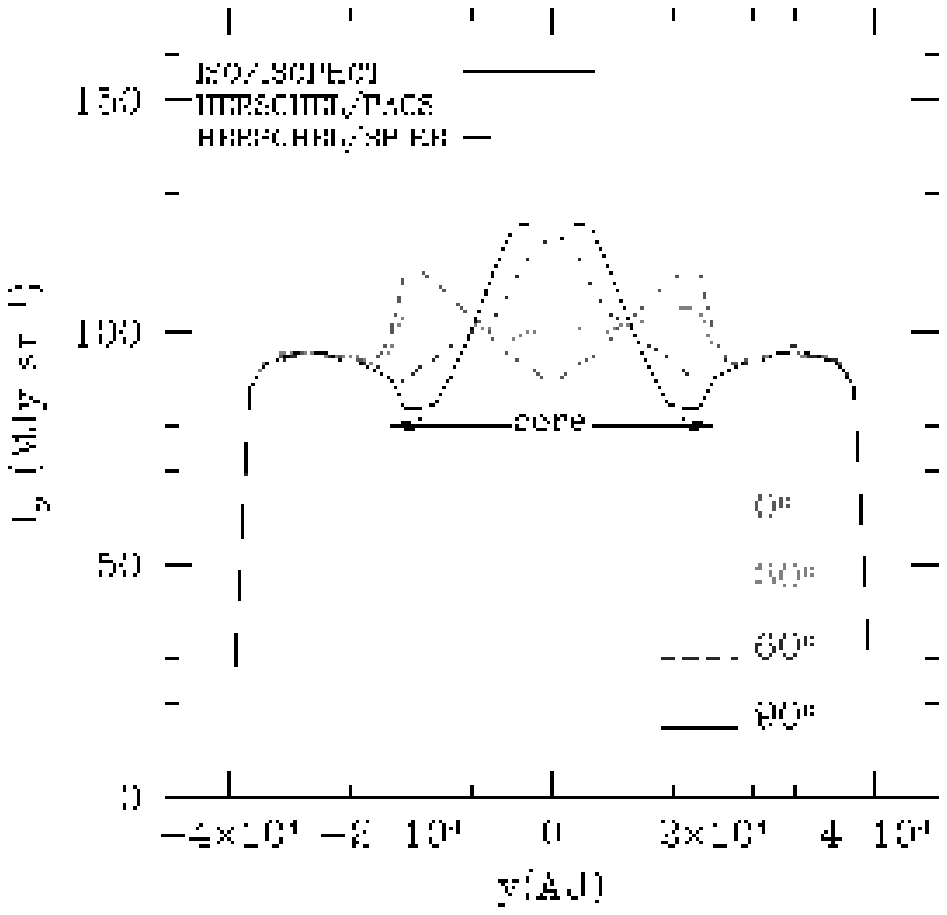}}
\caption{Same as  Fig.~\ref{image_rprof_nem}, i.e. model 1.2, but for a 
core embedded in a uniform molecular ambient cloud with visual extinction 
$A_{\rm V}=4$. The characteristic features visible at different viewing 
angles are weaker than in the case of a non-embedded core.}
\label{image_rprof_em}
\centerline{
\includegraphics[width=7cm]{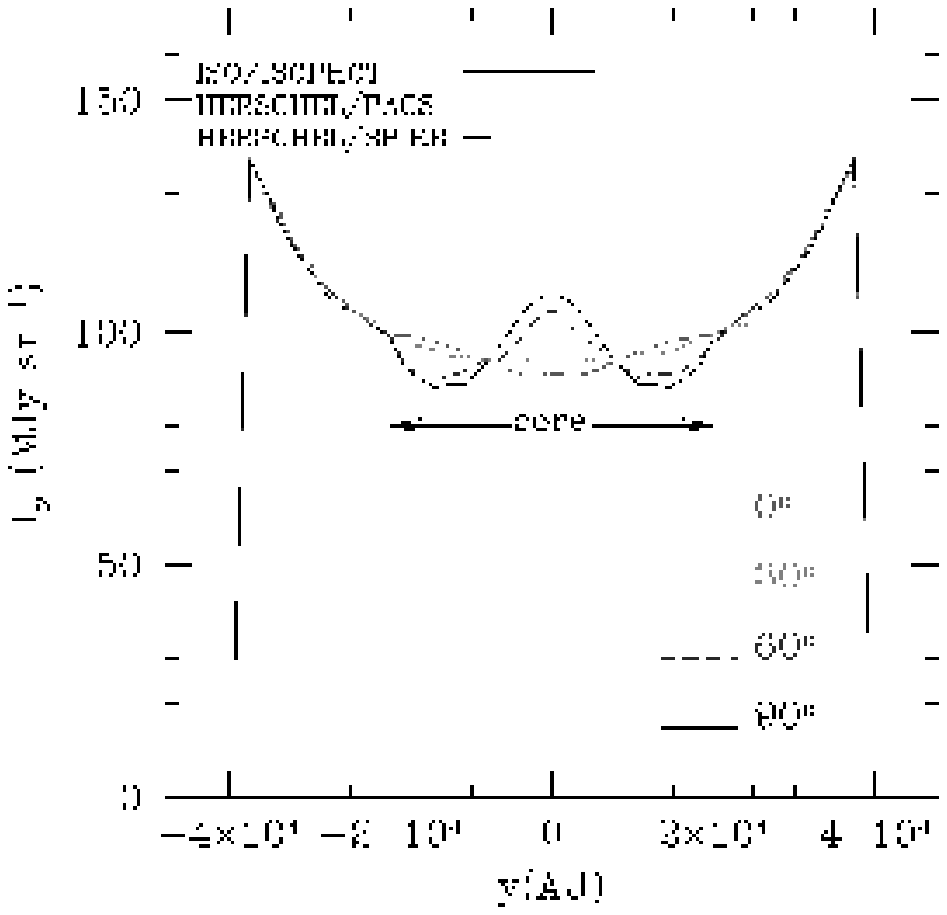}}
\caption{Same as  Fig.~\ref{image_rprof_nem}, i.e. model 1.2, but for a 
core embedded in a uniform molecular ambient cloud with visual extinction 
$A_{\rm V}=13$. There are two characteristic intensity minima visible at 
viewing angles 60$^{\rm o}$ and 90$^{\rm o}$, but at viewing angles 
0$^{\rm o}$ and 30$^{\rm o}$ these features are essentially invisible.}
\label{image_rprof_nem2}
\end{figure}

Here, we examine the more general case of embedded flattened cores. We 
model a core with the same set of parameters as model 1.2 ($p=4$, $e=2.5$) 
but embedded in a uniform density ambient cloud with different visual 
extinctions $A_{\rm V}$ ($A_{\rm V}=1.086\ \tau_{\rm V}$). The ambient 
cloud is illuminated by the BISRF. In Fig.~\ref{fig_asym_a.temp.comp}, 
we present the temperature profiles at $\theta=0\degr$ (core pole; 
upper curves) and $\theta=90\degr$ (core equator; lower curves), (i) 
for a non-embedded core (dashed lines; model 1.2), (ii) for the same 
core embedded at the centre of an ambient cloud with $A_{\rm V}=4$ 
(dotted lines), and (iii) for the same core embedded at the centre of 
an ambient cloud with $A_{\rm V}=13$ (full lines).

Relative to an non-embedded core, a core embedded in an ambient cloud 
with $A_{\rm V} = 4$ is colder and has lower temperature gradient (cf. 
Figs 3b and 15). The isophotal maps are similar to those of the 
non-embedded core (cf. Figs. 7 and 11), but the characteristic features 
are less pronounced. This is because the temperature gradient inside 
the core is smaller when the core is embedded (see 
Fig.~\ref{fig_asym_a.temp.comp}). For example, at half the radius of 
the core ($r = 10^4$~AU) the temperature difference between the 
point at $\theta=0\degr$ and the point at $\theta=90\degr$, is $5-6$~K 
for the non-embedded core, but only $\sim 1.5$~K for the same core 
embedded in an $A_{\rm V}=4$ ambient cloud. In Fig.~\ref{image_rprof_em} 
we present a perpendicular cut through the centre of the embedded core 
image at viewing angle 30\degr. It is evident that the features are 
quite weak, but they have the same size as in the non-embedded core 
(Fig.~\ref{image_rprof_nem}), and they may be detectable with {\it 
Herschel}, given an estimated rms sensitivity better than $\sim 1-3$ 
MJy sr$^{-1}$ at 170-250~$\micron$ for clouds outside the Galactic 
plane (dependent on cirrus confusion).

For a core embedded in an ambient cloud with $A_{\rm V}=13$, the 
temperature differences between different parts of the core are even 
smaller ($\la 1$~K at $r = 10^4$~AU, see 
Fig.~\ref{fig_asym_a.temp.comp}), but characteristic features persist 
(e.g. two symmetric intensity minima at 60\degr and 90\degr, see 
Fig.\ref{image_rprof_nem2}). 

Thus, continuum observations near the peak of the core emission, can 
be used to obtain information about the core density and temperature 
structure and orientation, even  if the core is very embedded 
($A_{\rm V}\sim 10-20$).

\subsection{The effect of a UV-enhanced ISRF on embedded cores}

\hspace{1.2em} We now examine the effect that a UV-enhanced  ISRF has 
on the temperature profiles and isophotal maps of deeply embedded cores. 
We consider an ISRF that consists of the BISRF, plus an  additional 
component of diluted blackbody radiation from stars with $T_\star=6000$~K 
or $T_\star=10000$~K, that illuminates isotropically the ambient molecular 
cloud in which the core resides. We use a dilution parameter 
$\omega_\star=10^{-13}$, so that the total additional luminosity 
illuminating the ambient cloud is
\begin{equation}
L=\omega_\star\, 4\pi R_{\rm cloud}^2\,  \sigma T_\star^4,
\end{equation}
where $\sigma$ is the Stefan-Boltzmann constant. This additional flux 
enhances the radiation at $\lambda=0.4~\micron$ incident on the ambient 
cloud by a factor of $\sim 3$ for $T_\star=6000$~K, and $\sim 30$ for $T_\star=10000$~K, as compared with the standard BISRF.

\begin{figure}
\centerline{
\includegraphics[width=7cm]{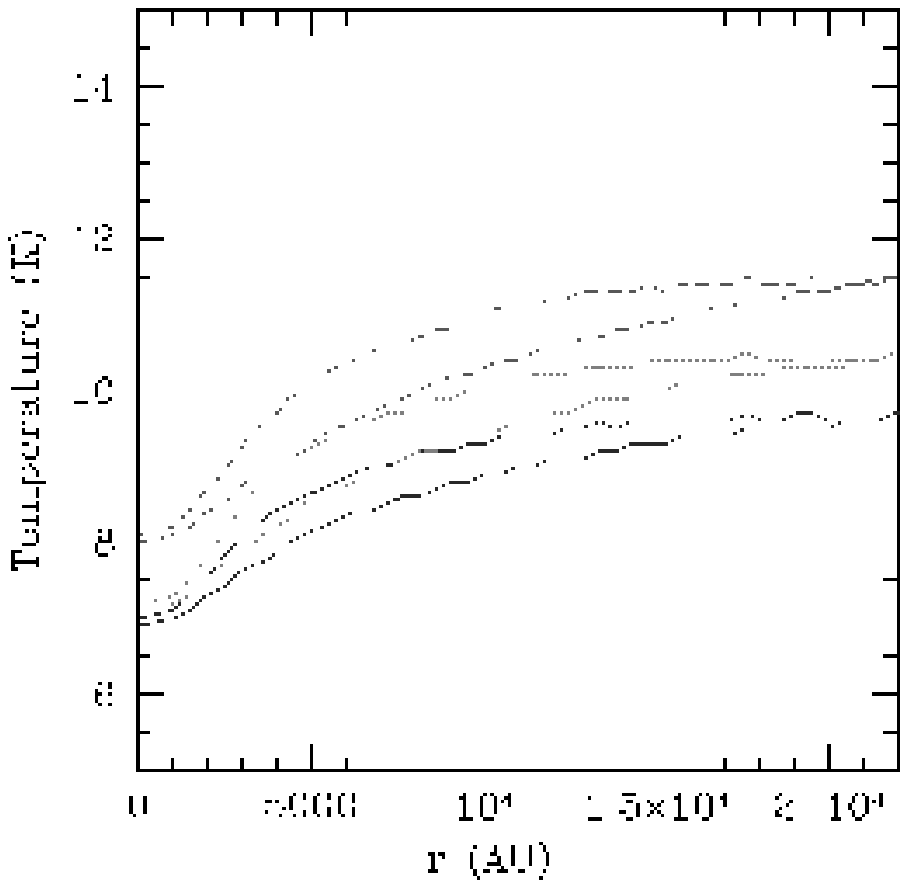}}
\caption{The effect of an UV-enhanced ISRF on embedded cores. Temperature 
profiles of a core with the same set of parameters as model 1.2 ($p=4$, 
$e=2.5$), embedded in a uniform density ambient cloud  with visual extinction 
$A_{\rm V}=13$, that is illuminated by the BISRF (solid lines), by the BISRF 
plus a diluted blackbody of  $T_\star=6000$~K (dotted lines) and by the BISRF 
plus a diluted blackbody of  $T_\star=10000$~K (dashed  lines). The upper 
curve of each set of lines corresponds to the direction towards the pole of 
the core ($\theta=0\degr$), and the bottom curve to the direction towards 
the core equator ($\theta=90\degr$). The core is hotter when the illuminating 
radiation field is enhanced at UV wavelengths, but the temperature differences 
between different parts of the core are not significantly increased.}
\label{fig_asym_a.temp.comp.uv}
\end{figure} 

In Fig.~\ref{fig_asym_a.temp.comp.uv}, we present the temperature profiles 
at $\theta=0\degr$ (core pole; upper curves) and $\theta=90\degr$ (core 
equator; lower curves) for a core with the same set of parameters as model 
1.2 ($p=4$, $e=2.5$), embedded in a uniform density ambient cloud  with visual 
extinction $A_{\rm V}=13$, illuminated by different ISRFs. The bottom pair 
of curves corresponds to illumination by the standard BISRF, and the upper 
two pairs of curves to illumination enhanced by a diluted blackbody with 
$T_\star=6000$~K and $T_\star=10000$~K. The core is  hotter when the ambient 
cloud is illuminated by a more energetic UV  field, by $1-2~K$ for the models 
we examine. The temperature differences between different parts of the core 
seem also to increase, but only by a small amount ($<0.5$~K). This means that 
the characteristic features on the isophotal maps at wavelengths near the peak 
of the core  emission are not significantly changed.

For an even more energetic illuminating radiation field the enhancement would 
be larger. For example, the external UV radiation field incident on $\rho$ 
Ophiuchi is estimated to be $\sim 10-100$ times stronger than the BISRF, 
due to the presence of a nearby B2V star (Liseau et al. 1999). In this 
circumstance, we would expect cores to be hotter and the temperature 
differences may then be sufficiently large ($\sim 2$~K) to produce detectable 
features on isophotal maps at 200~$\micron$. However, if the illuminating 
radiation field is very much stonger than the BISRF, it is likely to involve 
significant contributions from a few discrete luminous stars in the immediate vicinity. It will therefore be markedly anisotropic and this will produce 
additional asymmetries in the isophotal maps, making their analysis more 
difficult.

Thus, continuum observations near the peak of the core emission may reveal 
features characteristic of the core structure even if the core is more 
deeply embedded ($A_{\rm V}>20$), provided that the radiation field incident 
on the core is sufficiently intense, and provided the effect of discrete 
local sources can be treated.

\section{Axial asymmetry}
\label{sec:spole.cores}

\subsection{The model}

The density profile in model cores with axial asymmetry is given by
\begin{equation}
n(r,\theta) = 10^6\,{\rm cm}^{-3}\;\,\frac{1 + A \left( \frac{r}
{2000\,{\rm AU}} \right)^2 {\rm sin}^p(\theta/2) }{ \left[ 1 + 
\left( \frac{r}{2000\,{\rm AU}} \right)^2 \right]^2 } \,.
\end{equation}
Thus -- as with model cores having disk-like asymmetry -- the density 
is approximately uniform in the centre, and falls off as $r^{-2}$ in 
the outer envelope; the core has a spherical boundary at radius 
$R_{\rm core} = 2 \times 10^4\,{\rm AU}$; and the degree of asymmetry 
is determined by $A$ and $p$. The values of $A$ and $p$ we have treated 
are given in Table 2, along with
\begin{equation}
e = \frac{\tau_{\rm V}(\theta = 180\degr)}
{\tau_{\rm V}(\theta = 0\degr)} \,,
\end{equation}
$M_{\rm core}$ and $\tau_{\rm V}(\theta = 0\degr)$. $e$ is now the 
south-to-north pole optical depth ratio, i.e. the ratio of 
the maximum optical depth from the centre to the surface of the core 
(which occurs at $\theta = 180\degr$) to the minimum optical depth 
from the centre to the surface of the core (which occurs at $\theta = 
0\degr$). Isodensity contours on the $x=0$ plane are plotted on 
Figure 2 for the model with $e = 2.5$ and $p = 4$.

\begin{figure}
\centerline{
\includegraphics[width=6cm,angle=-90]{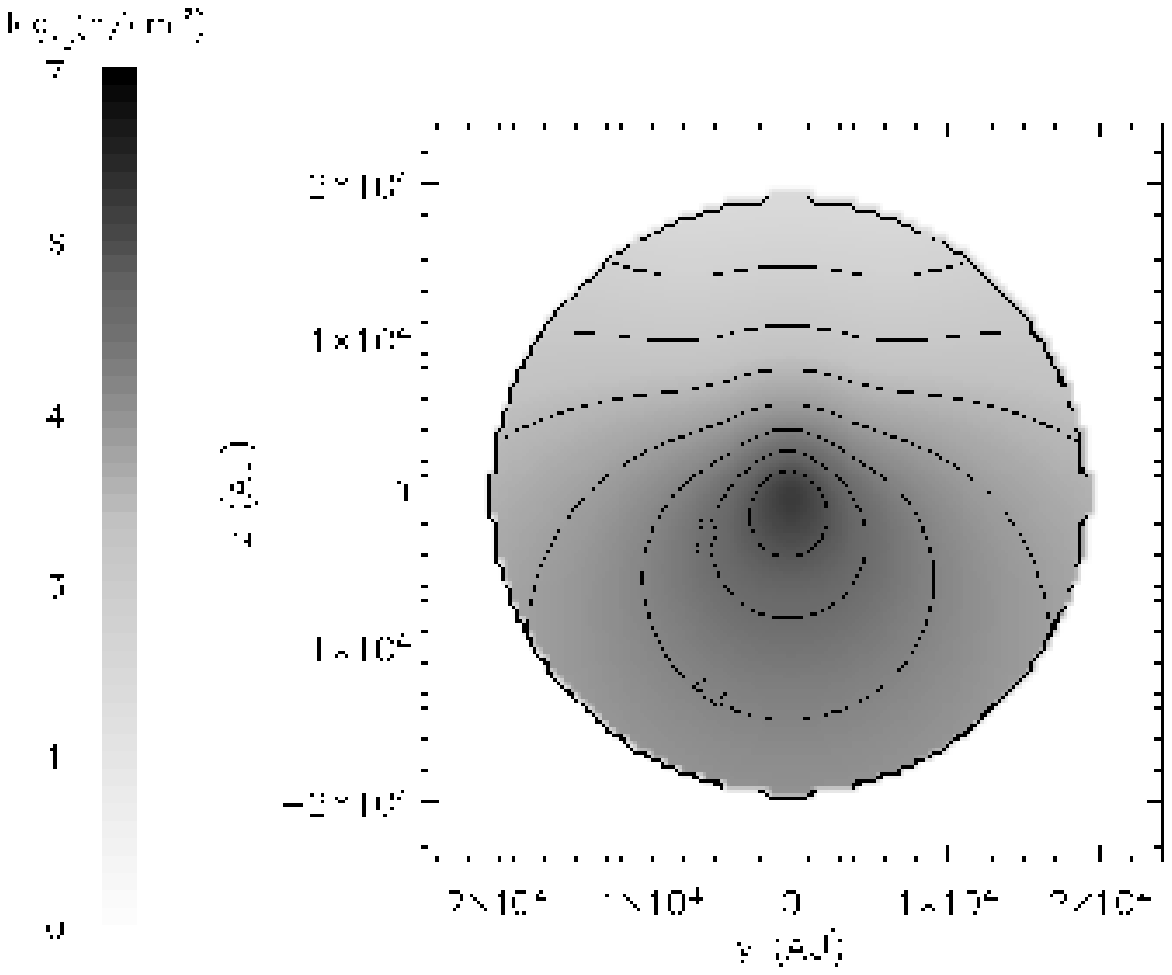}}
\caption{Density distribution on the $x=0$ plane for a core with 
axial asymmetry, south-to-north pole optical depth ratio $e=2.5$ 
and $p=4$ (model 2.2). We plot iso-density contours every $10^{0.5} 
{\rm cm^{-3}}$. The central contour corresponds to $n=10^{5.5} 
{\rm cm^{-3}}$.} 
\label{dens.asym.b.2.5}
\centerline{
\includegraphics[width=6cm,angle=-90]{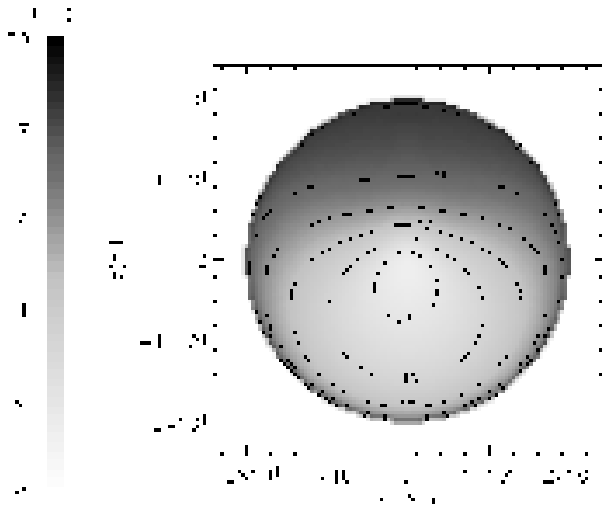}}
\caption{Temperature distribution on the $x=0$ plane, for the model 
presented in Fig.~\ref{dens.asym.b.2.5} ($e=2.5$, $p=4$, model 2.2). 
We plot iso-temperature contours from 8 to 18~K, every 2~K. The 
denser, southern parts of the core are colder.} 
\label{temp.asym.b.2.5}
\end{figure}

\begin{table}
\begin{center}
\caption{Model parameters for cores with axial asymmetry}
\begin{tabular}{@{}cccccc}
& & & & &\\
\hline
Model ID & $A$ & $p$ & $e$ & $M_{\rm core}/M_\odot$ & 
$\tau_{\rm V}(\theta = 0\degr)$ \\ \hline\hline
2.1 & 28 & 4 & 1.5 & 1.4 & 94 \\
2.2 & 81 & 4 & 2.5 & 3.4 & 94 \\
2.3 & 28 & 1 & 1.5 & 2.4 & 94 \\
2.4 & 81 & 1 & 2.5 & 6.3 & 94 \\
\hline
\hline
\end{tabular}
\begin{list}{}{}
\item[$e\;\;$:] South-to-north pole optical depth ratio
\item[$M_{\rm core}\;\;$:] Core mass
\item[$\tau_V(\theta = 0\degr)\;\;$:] Visual optical depth from the 
centre to the surface of the core along the pole ($\theta=0\degr$).
\end{list}
\label{tab:run_params}
\end{center}
\end{table}

\subsection{Results: core temperatures, SEDs and images}

In Fig.~\ref{dens.asym.b.2.5}, we present the core density profile on the 
$x=0$ plane for model 2.2 ($p=4$, $e=2.5$), and in  Fig.~\ref{temp.asym.b.2.5} 
the corresponding temperature profile. The temperature drops from $\sim 18$~K 
at the edge of the core to $\sim 7$~K at the centre of the core, and  the 
denser ``southern'' parts of the core are colder. The difference between 
regions of the core with the same $r$ but different $\theta$ is larger for 
the $p=4$ models than for the $p=1$ models, and also larger for the more 
asymmetric models ($e=2.5$) than for the less asymmetric models ($e=1.5$). 
For example, at half the radius of the core ($10^4$~AU) the 
temperature difference between the  point at $\theta=0\degr$ (core north 
pole) and the point at $\theta=180\degr$ (core south pole) is $\sim 7$~K for 
the $p=4$, $e=2.5$ model, $\sim 5$~K for the $p=4$, $e=1.5$ model, $\sim 4$~K 
for the $p=1$, $e=2.5$ model and $\sim 3$~K for the $p=1$, $e=1.5$ model.

\begin{figure}
\centerline{\includegraphics[width=5.7cm,angle=-90]{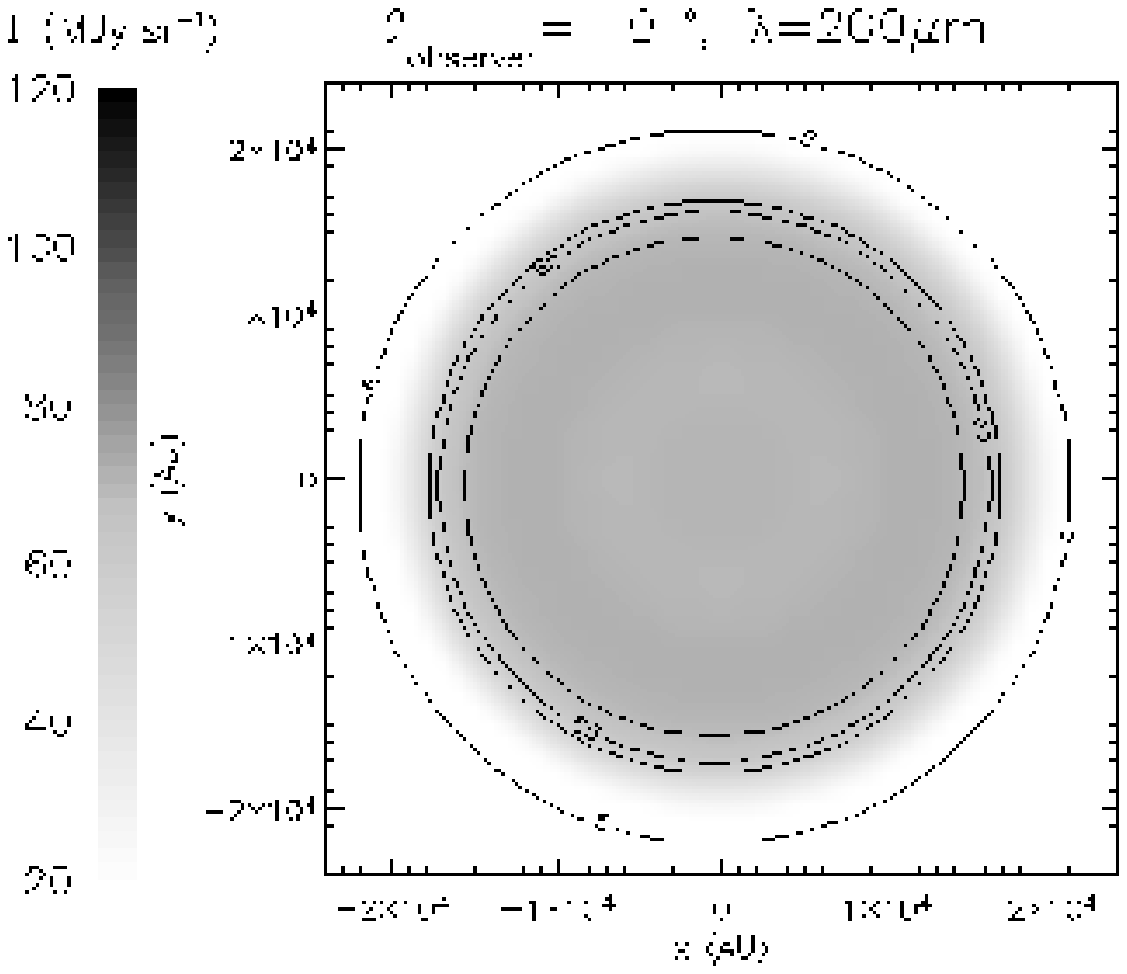}}
\centerline{\includegraphics[width=5.7cm,angle=-90]{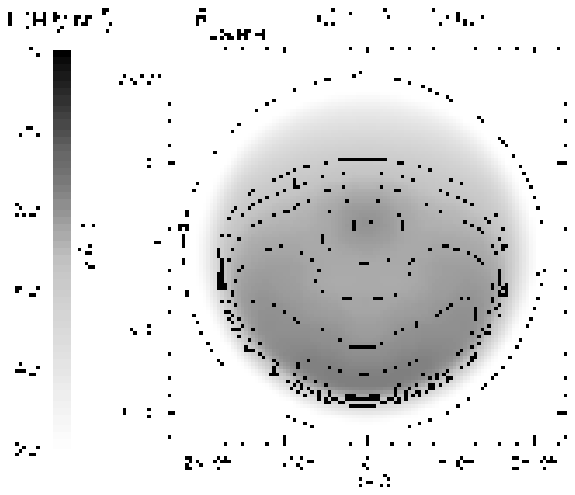}}
\centerline{\includegraphics[width=5.7cm,angle=-90]{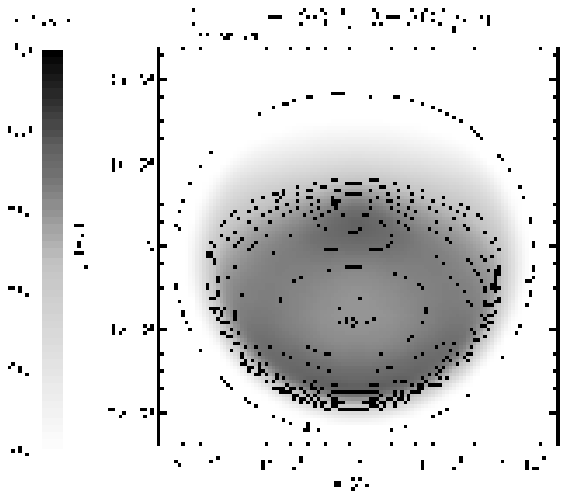}}
\centerline{\includegraphics[width=5.7cm,angle=-90]{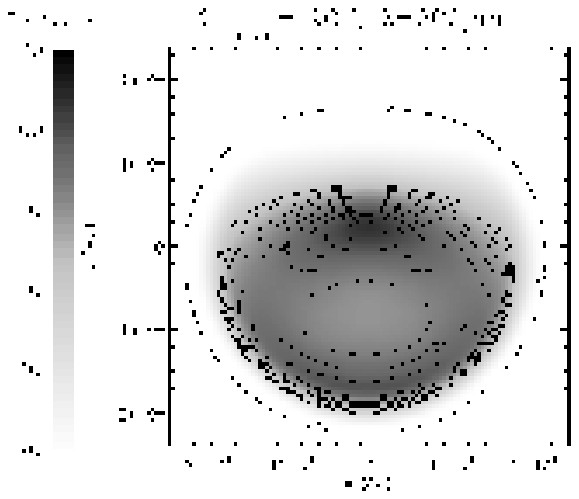}}
\caption{Isophotal maps at 200~$\micron$ at viewing angles $0\degr$, $30\degr$, 
$60\degr$ and $90\degr$, for a flattened core with south-to-north pole optical 
depth ratio $e=2.5$  and $p=4$ (model 2.2). We plot an isophotal contour at 
5~MJy~sr$^{-1}$ and then from 60 to 110~MJy~sr$^{-1}$, every 5~MJy~sr$^{-1}$.}
\label{images.asymb.2.5.200}
\end{figure}

\begin{figure}
\centerline{\includegraphics[width=5.7cm,angle=-90]{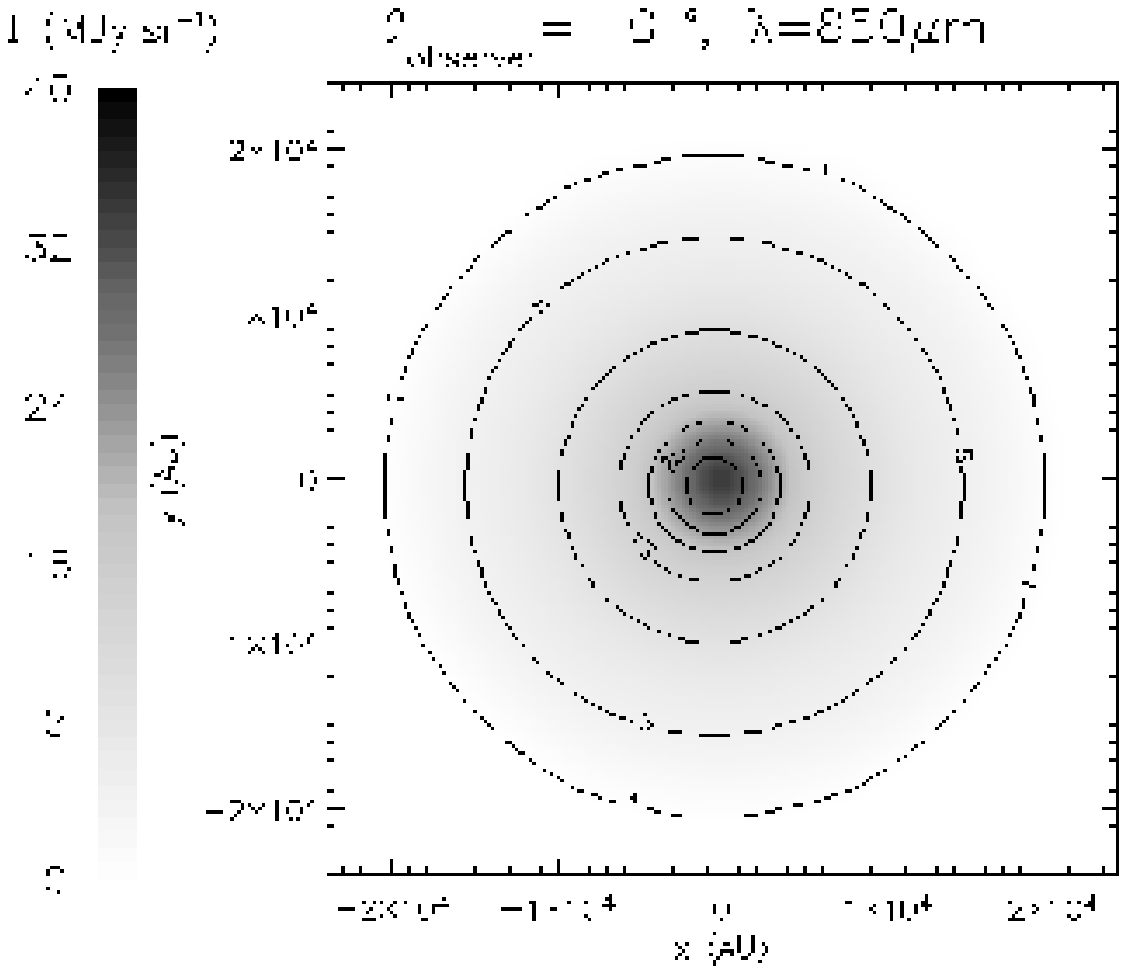}}
\centerline{\includegraphics[width=5.7cm,angle=-90]{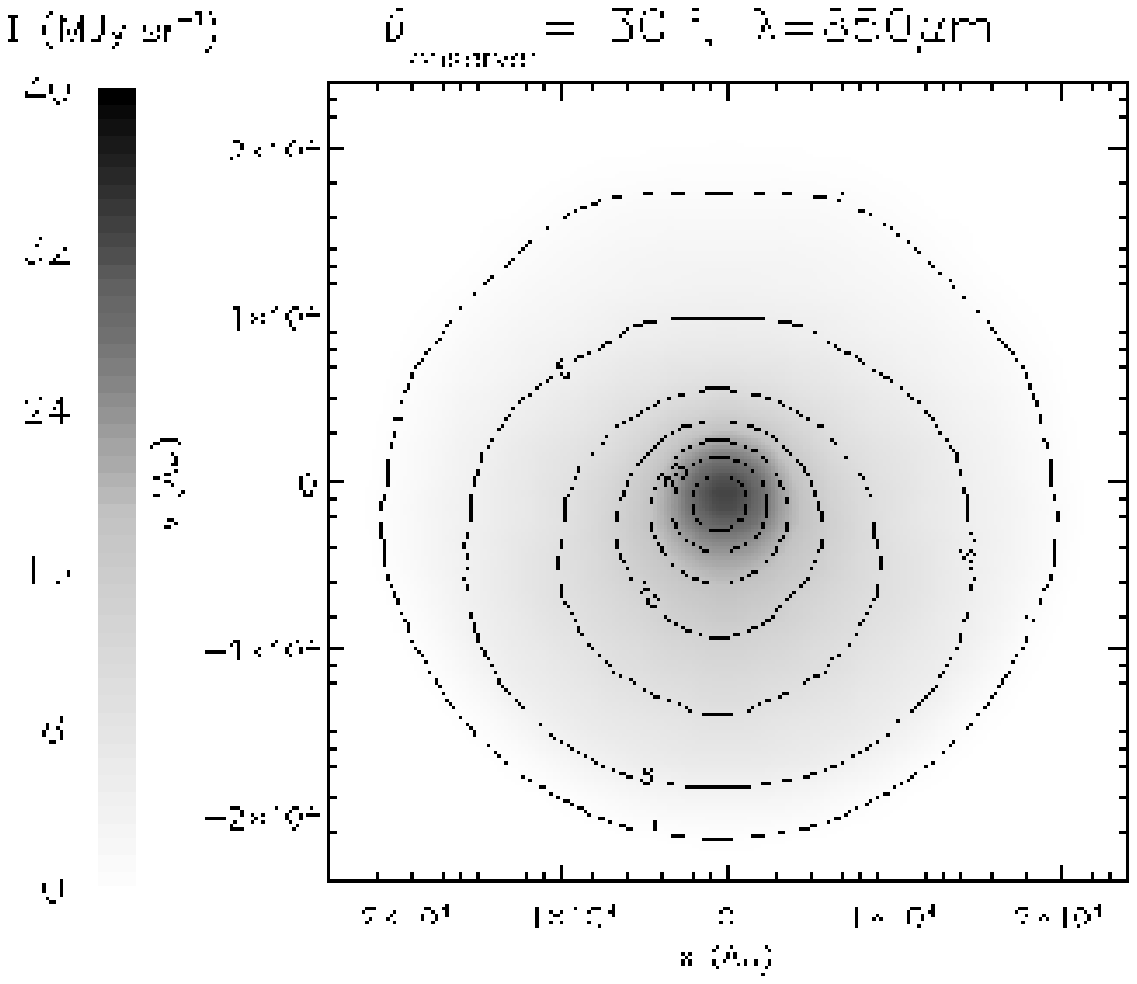}}
\centerline{\includegraphics[width=5.7cm,angle=-90]{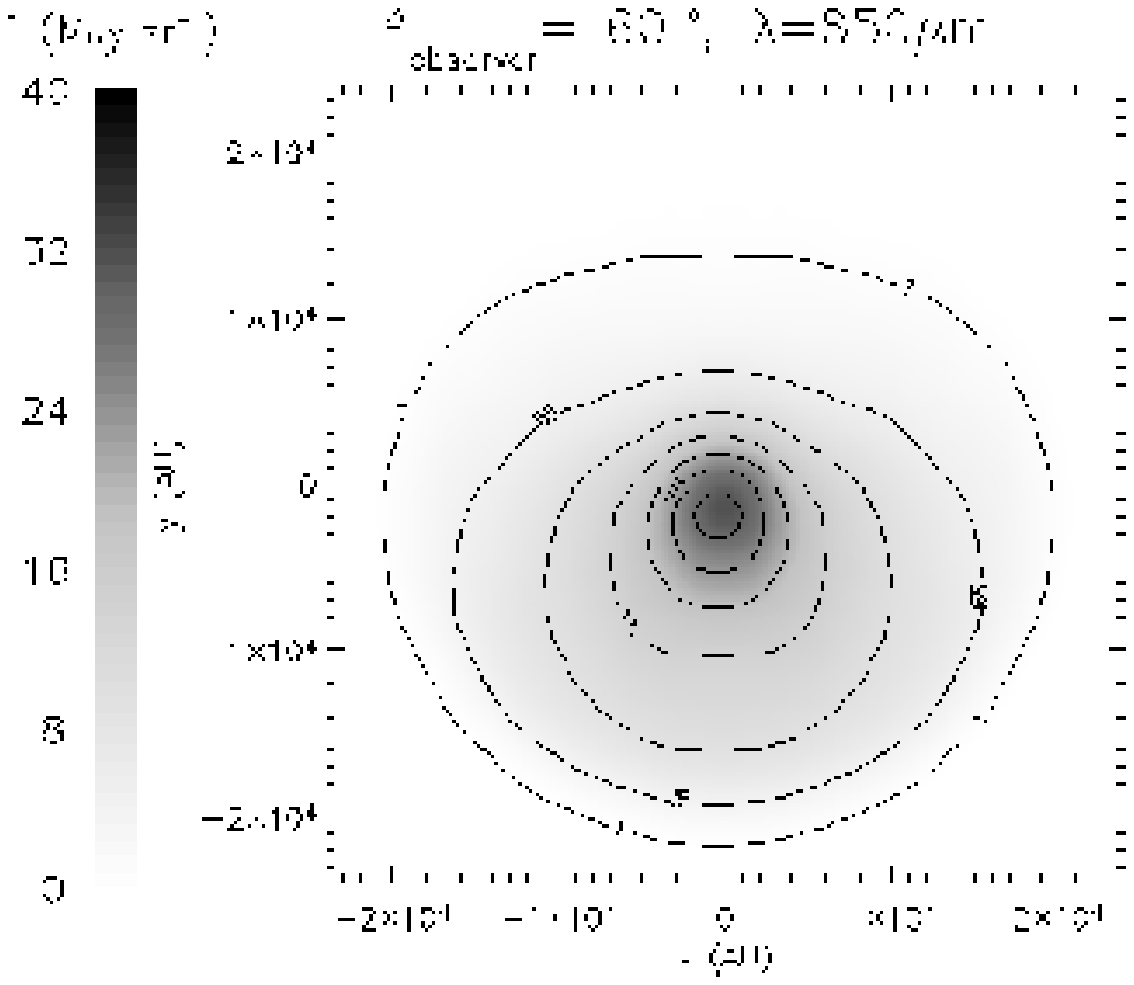}}
\centerline{\includegraphics[width=5.7cm,angle=-90]{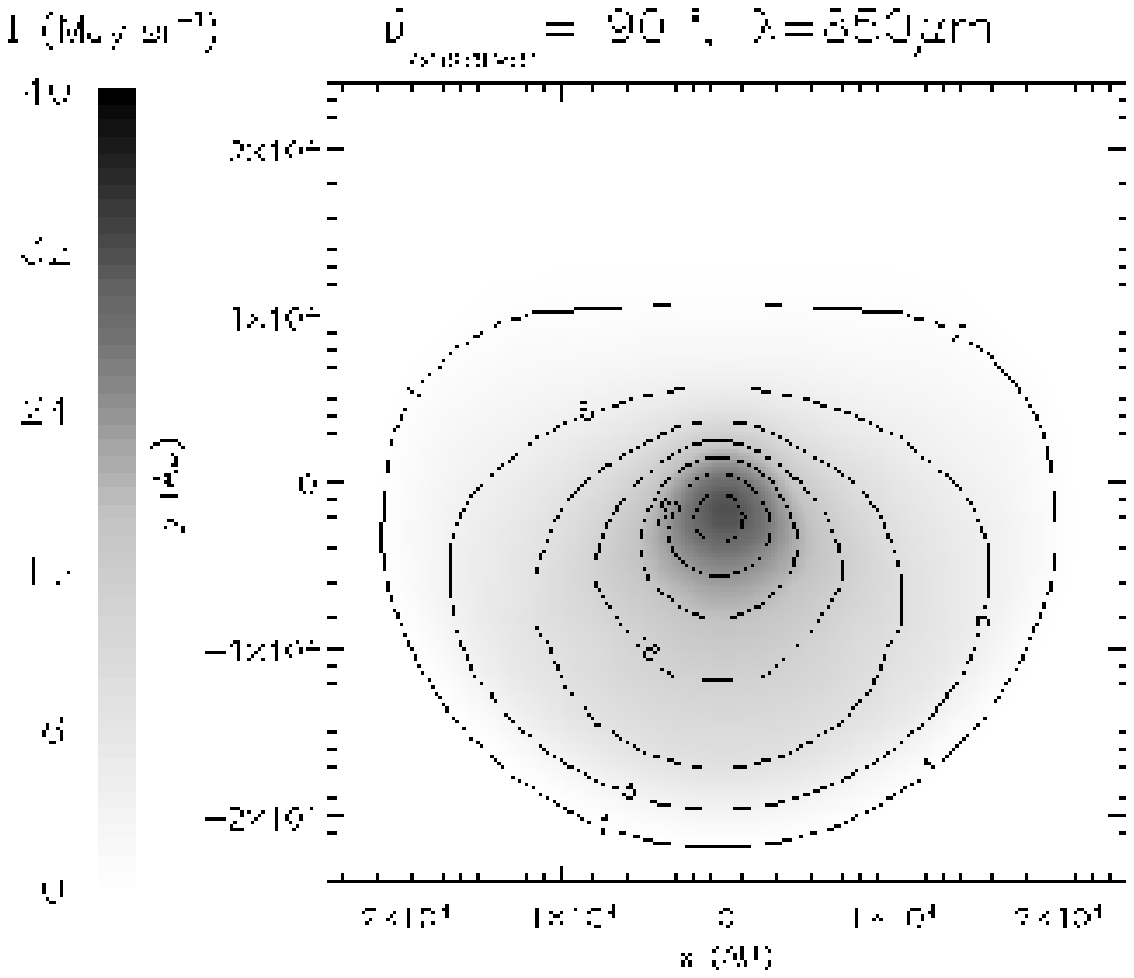}}
\caption{
Isophotal maps at 850~$\micron$ at viewing angles $0\degr$, $30\degr$, $60\degr$ 
and $90\degr$, for a flattened core with south-to-north pole optical depth ratio 
$e=2.5$ and $p=4$ (model 2.2). We plot an isophotal contour at 1~MJy~sr$^{-1}$ 
and then from 5 to 50~MJy~sr$^{-1}$, every 5~MJy~sr$^{-1}$. The core appears 
elongated when viewed at an angle other than $\theta=0\degr$.}
\label{images.asymb.2.5.850}
\end{figure}

As in the case of cores with disk-like asymmetry, these temperature 
differences result in 
characteristic features on isophotal maps at wavelengths near the peak of 
the core emission. In Fig.~\ref{images.asymb.2.5.200}, we present 
200~$\micron$ images at different viewing angles. The core appears 
spherically symmetric when viewed pole-on,  but the effects of axial 
asymmetry start to show when we look at the core from other viewing 
angles (e.g. $30\degr$, $60\degr$ and $90\degr$). Comparing with the 
images at 200~$\micron$ for cores with disk-like asymmetry 
(Fig.~\ref{images.asyma.1.5}-\ref{images.asyma2.2.5}), we see that for cores 
with axial asymmetry there is only one axis of symmetry. Thus, the symmetry of the 
characteristic features is indicative of the underlying core density structure. 
These features contain information about the core density, temperature 
and orientation with respect to the observer, and therefore observations 
near the peak of the core emission are important. The resolution of 
ISO/ISOPHOT was not high enough to detect such features.  

The 850~$\micron$ images (Fig.~\ref{images.asymb.2.5.850}) map the column 
density of the core along the line of sight. We point out the similarities 
between the maps in Fig.~\ref{images.asymb.2.5.850}, and SCUBA observations 
of L1521F, L1544, L1582A, L1517B, L63 and B133 (Kirk et al. 2004). Further 
modelling for each specific core is required to make more detailed comparisons. 

As in the case of cores with disk-like asymmetry, the SEDs of cores with 
axial asymmetry are independent of the observer's viewing angle, because 
they are optically thin at long wavelengths.

\section{Discussion}
\label{sec:asym.discussion}

We have performed accurate two dimensional continuum radiative transfer 
calculations for non-spherical prestellar cores. We argue that such 
non-spherical models are needed because observed cores are clearly not 
spherically symmetric, and are not expected to be spherically symmetric. 
Our models illustrate the characteristic features on isophotal maps 
which can help to constrain the intrinsic density and temperature fields 
within observed non-spherical cores. They demonstrate the importance 
of observing cores at wavelengths around the peak of the SED and with 
high resolution.

Our main results are:  

\begin{itemize}

\item For the cores treated here, which are optically thin at the long 
wavelengths where most of the emission occurs, the SED is essentially 
the same at any viewing angle. For example, in the case of cores having 
disk-like asymmetry, the SED does not distinguish between cores viewed 
from different angles.

\item Isophotal maps at submm wavelengths (e.g. at 850~$\micron$) are 
essentially column density tracers, whereas maps at far infrared wavelengths 
around the peak of the core emission (e.g. 200~$\micron$) reflect both the 
column density and the temperature field along the line of sight. Therefore, 
sensitive,  high-resolution observations at 170-250~$\micron$ ({\it Herschel}) 
combined with long-wavelength observations (e.g. 850~$\micron$ or $1.3~{\rm mm}$) 
can in principle be used to constrain the orientation of a core and the 
temperature field within it.

\item If we assume a universal ISRF, then cores embedded in ambient molecular 
clouds are colder than cores directly exposed to the ISRF and have lower 
temperature gradients within them. As a result, the characteristic features 
on 200~$\micron$ isophotal maps are weaker for cores embedded in ambient 
clouds with visual extinction less than $A_{\rm V}~\sim~10$, and may even 
disappear completely for more deeply embedded cores ($A_{\rm V}>20$).

\item If the ISRF incident on the ambient molecular cloud is enhanced in the UV 
region (for example, by nearby luminous stars) but still isotropic, then the 
embedded cores are hotter and the temperature gradients inside them may be 
sufficient to produce detectable characteristic features, even in deeply 
embedded cores.

\item The shapes of asymmetric cores depend strongly on the observer's 
viewing angle. For example, cores with disk-like asymmetry appear more 
flattened when viewed edge-on. Our models also indicate that such cores 
should be easier to detect when viewed from near the equatorial plane. 
This may introduce a selection effect that should be taken into account 
when studying the statistics of the shapes of cores, using solely optically 
thin continuum, or optically thin molecular-line, observations.
 
\item If the ambient ISRF is isotropic, the characteristic features on 
200~$\micron$ maps are symmetric with respect to two axes for cores with 
disk-like asymmetry, and with respect to one axis for cores with axial 
asymmetry. Thus, just the symmetry of these features, could be indicative 
of the core's internal density structure. Lack of symmetry in the 
features could indicate triaxiality, but it could also simply indicate that 
the radiation field incident on the core is anisotropic, due to discrete 
local sources.

\end{itemize}

Recently Gon\c{c}alves, Galli, \& Walmsley  (2004)
have presented radiative transfer models of axisymmetric and
non-axisymmetric toroidal cores, and also models of cores heated
by an additional external stellar source. In their approach, they use
a  similar but independent Monte Carlo radiative transfer code.
Their models focus on the effect of different core density profiles and
anisotropic heating of cores, whereas we focus on the effect of
observing mildly asymmetric cores at different viewing angles.
Both studies are in general agreement for similar core models, 
and are helpful for the study of non-spherical realistic prestellar cores.

We are now extending our study by treating the effect of an anisotropic 
illuminating radiation field. We are also studying the triaxial molecular cores 
that result from 3-dimensional hydrodynamic simulations.

\begin{acknowledgements}

We  gratefully acknowledge support from the EC Research Training Network
``The Formation and Evolution of Young Stellar Clusters'' (HPRN-CT-2000-00155).

\end{acknowledgements}

\end{document}